\documentclass[%
 aip,rsi,
 amsmath,amssymb,
reprint,%
]{revtex4-1}

\usepackage{graphicx}
\usepackage{dcolumn}
\usepackage{bm}
\usepackage{siunitx}
\begin{document}


\title{Characterising the optical response of ultra-low-noise far-infrared 60-110\,\bfseries{\si{\micro\metre}} transition edge sensors}

\author{E. A. Williams}
\author{S. Withington}
\author{D. J. Goldie}
\author{C. N. Thomas}
\affiliation{Cavendish Laboratory, University of Cambridge, J.J. Thomson Avenue, Cambridge CB3 OHE, United Kingdom}
\author{P. A. R. Ade}
\author{R. Sudiwala}
\affiliation{School of Physics and Astronomy, Cardiff University, Cardiff CF24 3YB, United Kingdom}

\date{\today}

\begin{abstract}

Far-infrared Transition Edge Sensors (TESs) are being developed for the SAFARI grating spectrometer on the cooled-aperture space telescope SPICA. In support of this work, we have devised a cryogenic (90 mK) test facility for carrying out precision optical measurements on ultra-low-noise TESs. Although our facility is suitable for the whole of the SAFARI wavelength range, 34-230\,\si{\micro\meter}, we focus on a representative set of measurements at 60-110\,\si{\micro\meter} using a device having a Noise Equivalent Power (NEP) of 0.32 $\mathrm{aW/\sqrt{Hz}}$. The system is able to perform a range of measurements: (i) Dark electrical characterisation. (ii) Optical efficiency with respect to a partially coherent beam having a modal composition identical to that of an ideal imaging telescope. (iii) Optical saturation and dynamic range. (iv) Fast optical transient response to a modulated thermal source. (v) Optical transient response in the presence of high-level background loading. We describe dark measurements to determine the operating characteristics of a TES,  and then compare predicted optical behaviour with measured optical behaviour. By comparing electrical and optical transient response, we were able to observe thermalisation in the device. We comment on the challenge of eliminating stray light.
\end{abstract}

\maketitle

\section{Introduction}

Transition Edge Sensors (TESs) use the strongly temperature-dependent resistance of a superconducting film biased on its transition to detect energy and power with exceptional sensitivity \cite{IrwinChapter}. Dominant in submillimetre-wave ground-based astronomy, TESs are under further development for the next generation of space observatories, including LiteBIRD \cite{matsumura2014mission,suzuki2018litebird},
SPICA \cite{roelfsema2018spica,audley2018safari,goldie2016performance,audley2014measurements,khosropanah2014characterization}, and
ATHENA \cite{barcons2017athena,jackson2016focal,gottardi2016development}, at submillimetre, infrared and X-ray wavelengths respectively.

The SAFARI instrument on SPICA is a grating spectrometer covering the wavelength range 34-230\,\si{\micro\meter} in three wavebands, each with a dedicated TES array. The complete instrument will require of order 2250 TESs read out using superconducting frequency domain multiplexers (FDM) \cite{jackson2011spica,den2012low,van2018active}. Because SAFARI will use a post-dispersed Martin-Puplett interferometer to provide both high and low resolution observing modes, the TESs will be arranged in three grating arrays of 5 $\times$ 150 pixels each. In addition, because the primary mirror of SPICA is cooled to 4 K, a Noise Equivalent Power (NEP) of 0.2 $\mathrm{aW/\sqrt{Hz}}$ is required to ensure that SAFARI is detector noise limited. This NEP is two orders of magnitude smaller than that needed for ground-based observatories, and can only be obtained by cooling the TES arrays, FDM filters and SQUID readout electronics to 50 mK.

Here we discuss the design and construction of a cryogenic test facility for characterising the dark and optical behaviour of ultra-low-noise TESs. The optical configuration enables optical efficiency to be determined with respect to a partially coherent beam having a modal composition identical to that of an ideal imaging telescope. In addition to steady-state illumination, a fast infrared thermal source allows TES transient response to tiny changes in optical power to be examined over timescales of tens of milliseconds. It is also possible to measure transient response in the presence of thermal background loading, increasing to the point of optical saturation. This information is important because the dynamical behaviour of the TES arrays must be matched to other time constants in the SAFARI instrument, such as the scan speed of the FTS, the slewing rate of the telescope, and readout and data rates.

Given the extreme sensitivity of these sensors, it was imperative to address stray light elimination, and magnetic and RF shielding, when configuring the apparatus. Ultimately, these considerations have direct implications for realising optimal TES performance in space-flight hardware. Here, we draw particular attention to some of the subtle problems associated with eliminating steady-state and fluctuating stray light.

We demonstrate TES characterisation using the optical test facility on a representative device in an array for the SAFARI M-band, 60-110\,\si{\micro\meter}. This is the first prototype SAFARI M-band array that we have fabricated, building on earlier demonstrations covering the 34-60\,\si{\micro\meter} and 110-210\,\si{\micro\meter} wavebands \cite{goldie2016performance,audley2014measurements,khosropanah2014characterization}. We report dark characterisation, TES optical efficiency, and electrical and optical transient response for parametric changes in steady-state background optical loading, the latter constituting our first such measurements on any TES device.

\section{Cryogenic optical test facility}
\label{subsec:EnhancedModule}

The test facility was based on a two-stage adiabatic demagnetisation refrigerator (ADR), backed by a pulse tube cooler (PTC). The PTC provided a 3.3\,K temperature stage, with the ADR giving an intermediate 1\,K stage and final 50\,mK stage. The ADR magnet current was controlled in a servo loop to hold the physical temperature of the TES array constant, at 90 mK unless otherwise specified in Section \ref{sec:Results}, with an RMS deviation of \SI{140}{\micro\kelvin} from target over a typical measurement set.

\begin{figure}
\centering
\includegraphics[]{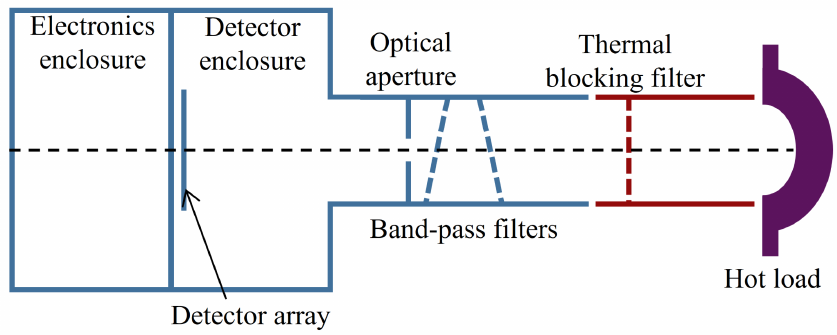}
\caption{\label{fig:SimplifiedOptical} Simplified representation of the test facility, showing the optical path from the blackbody hot load to the detector array.}
\end{figure}

A simplified representation of the optical configuration is shown in Fig. \ref{fig:SimplifiedOptical}: those components at the base temperature of 50\,mK are shown in blue, whereas those at 3.3\,K are in red. A cryogenic blackbody infrared load (shown in purple), with a slow response time, illuminated the detector array through a thermal radiation filter, a band-defining filter stack, and an optical aperture. The physical, and therefore radiometric, temperature of the hot load could be varied from 3.3\,K to 50\,K to enable the steady state optical response of the TES array to be measured.

The optical configuration has the following key features: (i) The far-field aperture of the test system and the input aperture of each detector ensure that the detector is illuminated by a specific number of optical modes, having the same spatial forms, and present in the same relative proportions, as those generated by an ideal imaging telescope. This feature is important because the detectors are partially coherent, and therefore their optical behaviour depends on the coherence properties of the illuminating field. (ii) The filters that determine the spectral band are placed ahead of the far-field aperture to ensure their presence does not significantly alter the spatial forms of the illuminating modes at the detector. In this way, the efficiency of a detector could be measured with respect to a well-defined partially coherent few-mode field.

\begin{figure*}
\centering
\includegraphics[]{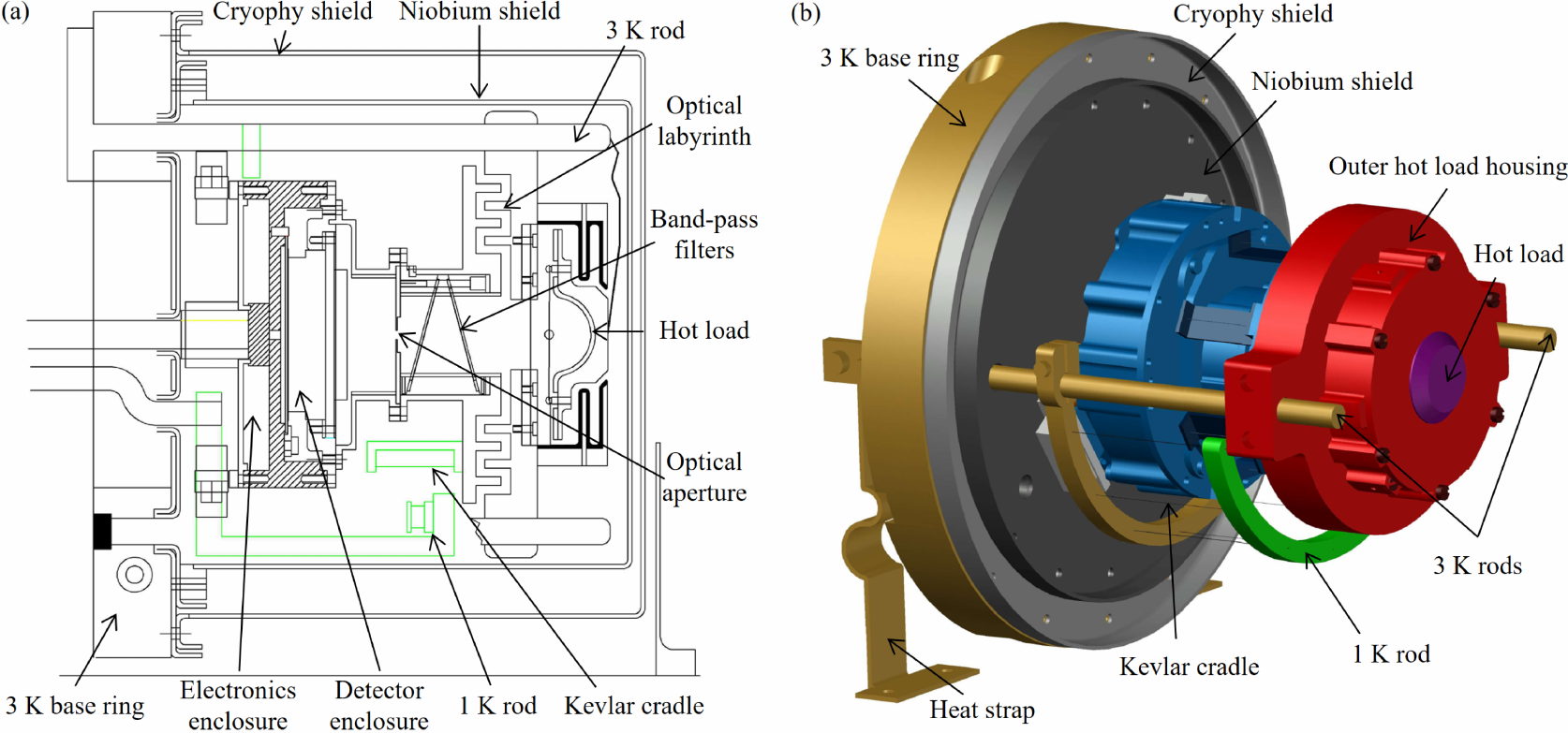}
\caption{\label{fig:NewModuleSchem} Schematics of the optical test facility design: (a) cross-section; (b) 3-D rendering without enclosing shields.}
\end{figure*}

The detector array was mounted in a light-tight enclosure, such that the detectors could be coupled to the incoming radiation through an array of micro-machined infrared lightpipes, or opened up to the beam to allow direct illumination. Direct illumination enables the performance of the lightpipes to be separated from the intrinsic performance of the detectors. An additional recess was machined into the back of the light-tight enclosure to house the bias electronics and SQUID readout chips. 

A detailed schematic of the test facility is shown in Fig. \ref{fig:NewModuleSchem}(a), and rendered in three dimensions in Fig. \ref{fig:NewModuleSchem}(b). The hot load, filters, aperture, and detector and electronics enclosure, indicated in Fig. \ref{fig:SimplifiedOptical}, may be identified. The optical components at 3 K were mounted on two rods to enable rapid assembly and disassembly.

The blackbody hot load consisted of a copper cone, coated on its inner face with a high emissivity SiC and C loaded epoxy. Kevlar threads were used to suspend the cone from the 3.3\,K level, and heating resistors and a cryogenic thermometer were used to produce a thermal radiation field having a controllable temperature. Crucially, the response time of the hot load was determined by connecting a wire between the Cu cone and one of the 3.3\,K rods, enabling the temperature of the hot load to be swept quickly without incurring an excessive thermal load. A thermal blocking filter was mounted on the 3.3\,K stage to reduce long wavelength heating of the band-defining filters, 60-110\,\si{\micro\meter}, which were held at 50\,mK. These custom-designed infrared high-pass and low-pass filters, made of patterned metallic films on polypropylene substrates \cite{ade2006review}, were angled and mounted in a blackened tube to avoid multiply scattered radiation reaching the detectors. As seen in Fig. \ref{fig:SimplifiedOptical}, a thermal break was necessary between the hot load and thermal filter, and the filter stack and detector. Simply leaving a gap risked allowing stray light to enter the light path from the environment of the cryostat, whose radiometric temperature is poorly defined and could be high. A non-touching optical baffle was designed, and tested by placing a hot load outside of the path but inside of the cryostat. Radiation from this external load could not be detected by the TESs, despite their sensitivity, demonstrating that the optical baffle worked exceedingly well.

The main blackbody load allowed the TES array to be radiated with a variable temperature, 3.3\,K to 50\,K, few-mode radiation field. Because of its mass, and the need to limit the heat load on the ADR, the thermal time constant of the load was adjusted such that recovery to base temperature following a temperature increase of 8 K, of the order required for TES saturation in this configuration, could be achieved in approximately 20 s. This arrangement provided a steady state optical source, in the sense that it was significantly slower than the time constants of the TESs. A key requirement, however, was to be able to measure the transient optical response for different values of background optical loading. This was achieved by placing a tiny ceramic resistor, 1 mm$^3$, on the periphery of the optical beam between the hot load and the thermal filter. The cross section of the resistor was sufficiently small that the blockage of power from the main hot load was negligible. At these wavelengths, the resistor acts as a point source on the edge of the field of view of the detectors. The chip resistor was suspended by two copper wires attached to the 3 K housing using cryogenic epoxy. Given the small mass of the resistor, approximately 3 mg, and the diameter of the wire, 0.127 mm, an input electrical power of less than 25 mW was needed to achieve a calculated temperature increase to within 1 mK of 10 K in under 2 ms. Various current waveforms could be applied to the resistor, allowing different forms of modulation. This modulation scheme worked exceptionally well, enabling direct measurement of TES optical response to small changes in optical signal over a range of optical background loadings.

Considerable attention was dedicated to the design of the detector and electronics enclosures. Two recesses were machined into the front and back of a piece of Oxygen-free high thermal conductivity Cu rod. The front recess housed the TES chip, whereas the back recess housed the bias resistors and SQUIDs. By separating the detectors and readout electronics by a machined wall, we were able to prevent stray light from the readout electronics reaching the TESs, an effect that we had seen in previous design iterations. Additionally, the front and back lids had baffled edges to ensure that stray light could not enter from the environment of the cryostat. Previous experience had indicated that it is exceedingly difficult to make removable lids having extremely low light leakage. All of the inner surfaces of the assembly were coated with infrared absorber, and great care was taken with feedthrough wiring.

Precise metrology was needed to ensure that the detector and backing plate wafers were aligned laterally with respect to each other and the exit apertures of the lightpipes. The challenge was made even more demanding by the fact that the distance between the TES absorbers and the exit apertures of the lightpipes was just \SI{20}{\micro\meter}, and that between the absorbers and backshorts \SI{21.25}{\micro\meter}, along the whole length of the chip. The array was aligned laterally by four dowel pins, and constrained by a further four dowel pins, positioned to account for differential thermal expansion between the copper base plate and the silicon detector assembly. The array was secured by G10 fibreglass clamps fastened to raised bosses machined directly into the enclosure base plate. Precision metrology was carried out using a coordinate measuring machine and surface profiler to verify the positions of the dowel pins, the flatness of the base plate and the alignment and orientation of the detector array with respect to the backshort array. Crucially, the positions were measured after multiple cooling cycles, indicating that there was no lateral creep. The bias resistors and SQUIDs were mounted in the back of the detector module. To improve heat sinking and reduce the emission of thermal radiation, the bias resistors and SQUID chips were clamped directly onto the copper base plate through apertures in the fibreglass circuit board.

Given the extreme sensitivity of the detectors, they must be shielded from stray light, stray magnetic fields, electromagnetic interference, and microphonic pick-up. Of these, it proved particularly challenging to prevent stray light entering the test module, and indeed penetrating the TES enclosure. In a previous iteration of the test facility \cite{williams2018ultra}, where the hot load was thermally connected to a cylindrical radiation shield surrounding the 50\,mK components, we detected significant stray light, and proved by extensive measurements that this radiation was thermal in origin, had a spectrum longward of 2\,mm, and entered the TES enclosure through a path that did not involve the lightpipes. Despite these investigations, the exact source of this radiation and its route to the TES array was never conclusively identified.

To design the magnetic shielding, detailed finite element modelling was carried out. The final design comprised a nested arrangement of an e-beam welded niobium inner can \cite{SMF}, and a high-permeability alloy Cryophy\textsuperscript{\textregistered} outer can \cite{MagneticShields}. Re-entrant flanges were used on the edges of the cans to prevent stray fields entering through joints; these are detailed in Fig. \ref{fig:NewModuleSchem}(a). Finally, to minimise flux leakage,  openings in the shields were restricted to those in the face adjoining the 3 K copper base ring on the left of Fig. \ref{fig:NewModuleSchem}(a). Overall, the flux attenuation factor with respect to the inner volume of the main cryostat was calculated to be of order $10^2$ for DC and $10^5$ for AC external fields of amplitude 40\,nT. 

To reduce microphonic noise, great care was taken to prevent loose wiring sweeping through stray magnetic fields in response to mechanical vibrations. Wiring passing from the 3\,K rod to the 1\,K rod was therefore secured to Kevlar straps running between the 50\,mK, 1\,K and 3\,K stages. The chosen wiring scheme used twisted pair Nb/Ti in a CuNi matrix for optimal thermal isolation, with all non-twisted-pair wire loops minimised.

\begin{figure}
\centering
\includegraphics[]{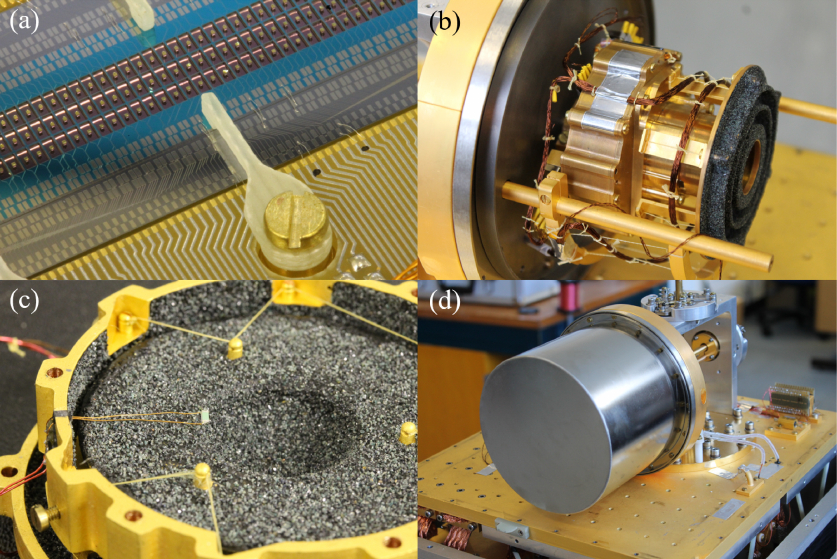}
\caption{\label{fig:NewModuleImages} Images of the test module through various stages of assembly. (a) Detector enclosure with mounted TES array. The TES array is seen in blue, and the Si backing plate with micromachined Au plated backshorts is seen in grey.
(b) Installed 50 mK sections including the electronics and detector enclosures, enclosed filters and optical labyrinth. (c) Hot load prior to installation. The optical modulator can just be seen as a white chip suspended on fine Cu wires. (d) Assembled module viewed from the exterior of the magnetic shields, above the 3 K base plate.}
\end{figure}

Figure \ref{fig:NewModuleImages} shows images of the test facility at various stages of assembly. Figure \ref{fig:NewModuleImages}(a) shows the TES array clamped within the detector enclosure, and superconducting fan-out wiring. Installed 50 mK sections are shown in Fig. \ref{fig:NewModuleImages}(b) prior to the hot load being mounted, including the electronics and detector enclosures, optical filter housing, and the 50 mK section of the main optical labyrinth. The 3 K rods can be seen, with wiring secured to a Kevlar cradle between the different temperature stages. Figure \ref{fig:NewModuleImages}(c) shows the hot load prior to installation, suspended from its outer housing by Kevlar threads. The tiny optical modulator is also visible. Finally, the module is viewed from the exterior of the magnetic shielding in Fig. \ref{fig:NewModuleImages}(d).

\section{M-Band TES Array}
\label{subsec:TESs}

\begin{figure}
\centering
\includegraphics[]{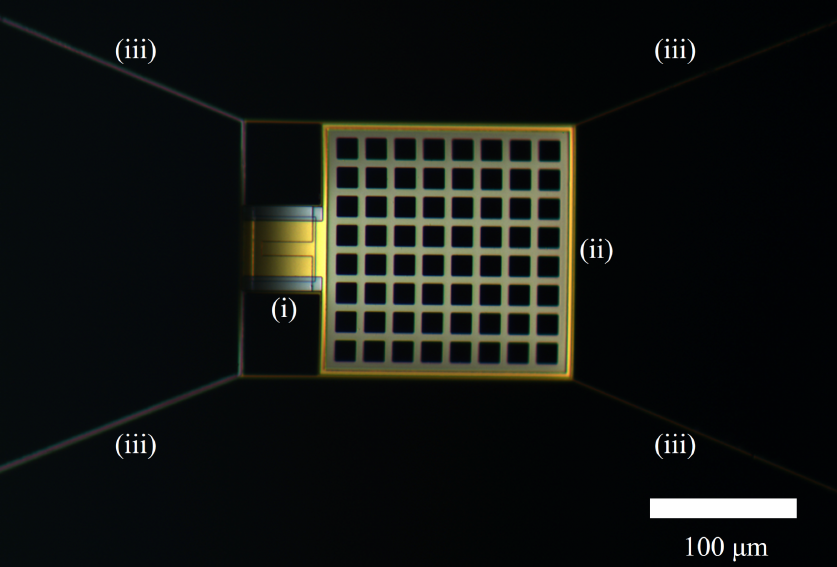}
\caption{\label{fig:TESImage} Optical microscopy image of the tested ultra-low-noise FIR TES, with superconducting Mo/Au bilayer (i), meshed $\beta$-Ta absorber with Au thermalisation ring (ii) and four $\mathrm{SiN_x}$ support legs (iii).}
\end{figure}

TESs were fabricated on a 200 nm thick, amorphous, low-stress $\mathrm{SiN_x}$ membrane, in a 3$\times$37 linear geometry suitable for grating spectrometer readout \cite{glowacka2012fabrication}. The array is shown installed in the test facility in Fig. \ref{fig:NewModuleImages}(a).
The results described in this paper were based on the representative TES shown in Fig. \ref{fig:TESImage}.

Each TES consisted of a 50$\times$50\,\si{\micro\meter} Mo/Au superconducting bilayer, Fig. \ref{fig:TESImage}(i), thermally coupled to a 170$\times$170\,\si{\micro\meter} $\beta$-phase Ta FIR absorber (ii), forming an island suspended from the surrounding wafer by four SiN$_\mathrm{x}$ support legs (iii) of length \SI{640}{\micro\metre} and cross-section $0.2\times1.5$\,\si{\micro\metre}. Nb wiring was deposited on two of the legs for biasing and readout. Varying numbers of interdigitated Au bars were deposited on the upper surfaces of the bilayers, giving transition temperatures $T_\mathrm{C} = 139 \pm 6$ mK. Each TES was fabricated with either a solid absorber, or a meshed absorber consisting of grids of square \SI{15}{\micro\metre} side-length apertures penetrating both the Ta film and the underlying SiN$_\mathrm{x}$ membrane, with a 47$\%$ filling factor. Since the response time of a TES is fundamentally related to the heat capacities of its components, the meshed design was developed to reduce the heat capacity associated with the absorber. The Ta film thickness was increased to 17 nm from the 8 nm of a solid absorber, to obtain an effective surface impedance matching free space.

To facilitate maximum power absorption by the absorber, an optically flat reflective backshort was placed at a distance of $\lambda_\mathrm{C}$/4 behind each absorber, where $\lambda_\mathrm{C}$ is the band-centre wavelength taken to be \SI{85}{\micro\metre} for the SAFARI M-band. The backshorts consisted of a high-conductivity sputtered Au film on an array of Si pillars, etched from the upper layer of a silicon-on-insulator (SoI) backing wafer, whilst also defining a recess into which the TES array chip was mounted. Nb breakout wiring was deposited on the backing plate.

Under normal operation, the TESs would be illuminated by an array of micromachined lightpipes, mounted directly above the detector array. Tapered pyramidal lightpipes were manufactured having walls of thickness 150\,\si{\micro\metre}, \SI{1350}{\micro\metre}\,$\times$\,\SI{650}{\micro\metre} entrance apertures and \SI{120}{\micro\metre}\,$\times$\,\SI{120}{\micro\metre} square exit apertures. The lightpipe array was then mounted such that the exit apertures were aligned axially with the corresponding \SI{170}{\micro m^2} TES absorbers. We have previously installed and tested the lightpipe array with the M-band TES array \cite{williams2018ultra}. The lightpipes were removed for the measurements described in this paper however, so that the planar TES absorber could be illuminated directly, allowing the TES optical efficiency to be determined independently of the modal transfer properties of the lightpipes.

The device under test was voltage-biased with a low impedance source defined by a 1.45 m$\Omega$ bias resistor and read out using a two-stage SQUID amplifier as a low-noise current-to-voltage converter. The SQUIDs were manufactured by Physikalisch-Technische Bundesanstalt \cite{drung2007highly}. A key requirement for maintaining stable voltage bias of the low-impedance TESs is to keep the total stray resistance below \SI{0.1}{m\ohm}.  To ensure this, superconducting wiring was used for all connections between the voltage source, SQUID input coil and TES, including bond wires and traces on the chip and readout circuit board.

\section{TES electrothermal modelling}
\label{sec:ElectrothermalMod}
\setcounter{equation}{0}

One aspect of TES characterisation using the optical test facility is the measurement of optical and electrical response times. Here we briefly describe the theoretical model that was used for interpreting the experimental data described in Section \ref{subsec:Risetimes}.

\begin{figure}
\centering
\includegraphics[]{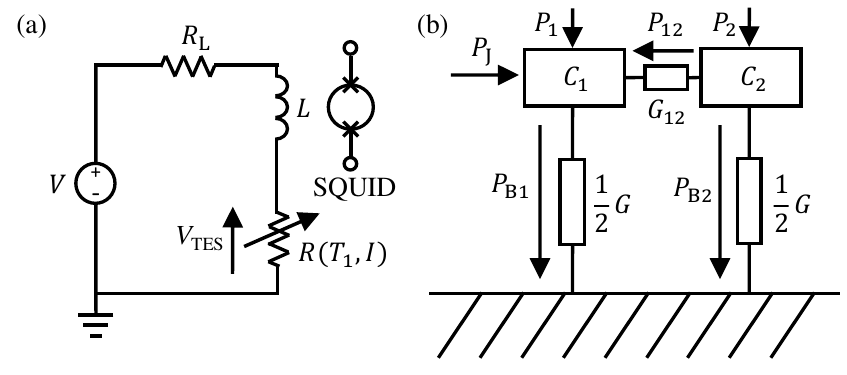}
\caption{\label{fig:Circuits} (a) Th\'evenin equivalent representation of the TES bias circuit, where the TES is shown as a variable resistor
$R(T_1, I)$. $R_\mathrm{L}$ is the internal resistance of the voltage source, $V$, and is the sum of a 1.45 m$\Omega$ bias resistor and an $\approx 0.5$ m$\Omega$ stray resistance. $L$ corresponds to the input inductance of the SQUID and any stray wiring inductance. $V_\mathrm{TES}$ is the voltage across the TES. (b): Thermal circuit representing the TES as two heat capacities, $C_1$ and $C_2$, where $C_1$ contains the superconducting bilayer, connected by a weak thermal link with conductance $G_{12}$. Each capacity is also coupled to the heat bath via thermal conductance $G/2$ where $G$ is the combined thermal conductance of the TES legs. $P_\mathrm{B1}$ and $P_\mathrm{B2}$ represent net thermal power to the heat bath from $C_1$ and $C_2$ respectively, $P_{12}$ is net thermal power between $C_1$ and $C_2$, $P_\mathrm{J}$ is Joule power input to $C_1$, and $P_1$ and $P_2$ are external power inputs to $C_1$ and $C_2$.}
\end{figure}

A voltage-biased TES self-regulates its temperature to within a narrow range around the transition temperature. Absorbed optical power causes the temperature of the bilayer to increase, increasing its resistance. However, this in turn causes the dissipated Joule power to decrease, maintaining the operating point of the device. This negative feedback enhances many aspects of performance, including speeding up the device with respect to its intrinsic thermal response. Figure \ref{fig:Circuits}(a) shows a Th\'evenin representation of the voltage-bias and readout circuits. The TES is represented as a variable resistor with temperature- and current-dependent resistance $R(T_1, I)$, where $T_1$ is the bilayer temperature and $I$ is the current. $V$ is the Th\'evenin-equivalent bias voltage and $V_\mathrm{TES}$ is the voltage across the TES. The current is read out using an inductively coupled SQUID. $R_\mathrm{L}$ is the sum of the bias and stray resistances, and $L$ represents the sum of the input inductance to the SQUID and any additional stray inductance.

To represent a TES bolometer with a large optical absorber adjacent to the superconducting bilayer, a simple thermal model has been adopted featuring two heat capacities, $C_1$ and $C_2$, thermally linked with conductance $G_{12}$. Each heat capacity is also connected to the surrounding heat bath with conductance $G/2$, where $G$ is the combined conductance of the TES legs, as illustrated in Fig. \ref{fig:Circuits}(b). The first heat capacity is taken to contain the bilayer, where Joule power is dissipated. The second heat capacity may be interpreted as being associated with the absorber and underlying dielectric. However, subject to the constraints implied by the choice of conductance to the heat bath, it is not necessary to define a physical location for this heat capacity in order to construct the model.

TES electrothermal behaviour is described by the following electrical and thermal differential equations:
\begin{align}
L\frac{\mathrm{d}I}{\mathrm{d}t} &= V-IR_\mathrm{L}-IR(T_1,I), \label{eq:ElectrothermalDiff1}\\
C_1\frac{\mathrm{d}T_1}{\mathrm{d}t} &= -P_\mathrm{B1}+P_{12}+P_\mathrm{J}+P_1, \label{eq:ElectrothermalDiff2}\\
C_2\frac{\mathrm{d}T_2}{\mathrm{d}t} &= -P_\mathrm{B2}-P_{12}+P_2, \label{eq:ElectrothermalDiff3}
\end{align}
where $T_1$ is the temperature of the first heat capacity, equal to the bilayer temperature, and $T_2$ is the temperature of the second heat capacity, equal to the absorber temperature. $P_\mathrm{J}=I^2R$ is the Joule power dissipated in the bilayer. $P_\mathrm{B1}$ and $P_\mathrm{B2}$ are thermal powers to the heat bath from each heat capacity; $P_{12}$ is the net power flow between the heat capacities; and $P_1$ and $P_2$ are additional power inputs, for example optical absorption, to the two heat capacities respectively. $R_\mathrm{L}$ is the internal resistance of the voltage source, $V$, and is the sum of a 1.45 m$\Omega$ bias resistor and an $\approx 0.5$ m$\Omega$ stray resistance. $L$ corresponds to the input inductance of the SQUID and any stray wiring inductance.

It is standard practice in TES physics to expand non-linear terms \cite{IrwinChapter}, in this case $P_\mathrm{B1}$, $P_\mathrm{B2}$, $P_{12}$, $P_\mathrm{J}$ and $R(T_1,I)$, to first order in the small signal limit around the initial steady state values $I_0$, $R_0$, $T_{10}$ and $T_{20}$, giving
\begin{equation}
\frac{\mathrm{d}}{\mathrm{d}t}
\begin{pmatrix}
\delta I \\
\delta T_1\\
\delta T_2\\
\end{pmatrix}
=-\mathbf{M}
\begin{pmatrix}
\delta I \\
\delta T_1\\
\delta T_2\\
\end{pmatrix}
+
\begin{pmatrix}
\frac{\delta V}{L} \\[0.5em]
\frac{\delta P_1}{C_1} \\[0.5em]
\frac{\delta P_2}{C_2}
\end{pmatrix},
\label{eq:Diff_Matrix}
\end{equation}
where
\begin{equation}
\mathbf{M}=
\begin{pmatrix}
\frac{R_\mathrm{L}+R_0(1+\beta)}{L} & \frac{I_0R_0\alpha}{T_{10}L} & 0 \\[0.7em]
-\frac{I_0R_0(2+\beta)}{C_1} & -\frac{\alpha \frac{I_0^2R_0}{T_{10}}-\frac{1}{2}G-G_{12}}{C_1} & -\frac{G_{12}}{C_1} \\[0.7em]
0 & -\frac{G_{12}}{C_2} & \frac{\frac{1}{2}G+G_{12}}{C_2}
\end{pmatrix},
\label{eq:Matrix}
\end{equation}
and $\delta I=I-I_0$, $\delta T_1 = T_1-T_{10}$,  $\delta T_2 = T_2-T_{20}$. The resistance-temperature and resistance-current sensitivities are given by $\alpha=(\partial \mathrm{ln}R/\partial \mathrm{ln}T_1)_\mathrm{I}$ and $\beta=(\partial \mathrm{ln}R/\partial \mathrm{ln}I)_{T_1}$ respectively. $\delta V$, $\delta P_1$ and $\delta P_2$ represent small changes in the applied bias voltage and external power input to the first and second heat capacities. This results in a solution for $\delta I (t)$, $\delta T_1 (t)$ and $\delta T_2 (t)$ of the form:

\begin{align}
\begin{pmatrix}
\delta I \\
\delta T_1\\
\delta T_2\\
\end{pmatrix}
&=\displaystyle\sum_{i=1}^{3} (\mathbf{a_\mathrm{i}}-\mathbf{a_\mathrm{i}}\mathrm{e}^{-t/\tau_\mathrm{i}}) \label{eq:RtSSMod1}\\
&=\displaystyle\sum_{i=1}^{3} (b_\mathrm{i}\mathbf{v_\mathrm{i}}-b_\mathrm{i}\mathbf{v_\mathrm{i}}\mathrm{e}^{-t/\tau_\mathrm{i}}), \label{eq:RtSSMod2}
\end{align}

where $\tau_\mathrm{i}=\lambda_\mathrm{i}^{-1}$, $\lambda_\mathrm{i}$ being the eigenvalues and $\mathbf{v_\mathrm{i}}$ the eigenvectors of the matrix $\mathbf{M}$, and

\begin{equation}
\begin{pmatrix}
b_1\\
b_2\\
b_3
\end{pmatrix}
=-(\lambda_1 \mathbf{v_1}\quad \lambda_2 \mathbf{v_2}\quad \lambda_3 \mathbf{v_3})^{-1}
\begin{pmatrix}
\frac{\delta V}{L} \\[0.5em]
\frac{\delta P_1}{C_1} \\[0.5em]
\frac{\delta P_2}{C_2}
\end{pmatrix}.
\label{eq:bSS}
\end{equation}

\section{Measurements}
\label{sec:Results}

\subsection{Preliminary thermal and electrical characterisation}

On cooling without temperature regulation, the temperature of the electronics and detector enclosures reached $T_\mathrm{B}=59.5$ mK, and remained below 90 mK for 2.5 hours, confirming the general thermal design.

In order to verify the basic performance of the TES bias circuit, the stray resistance, $R_\mathrm{stray}$, and inductance, $L$, were extracted from the measured circuit impedance with the TES in its superconducting state \cite{lindeman2004impedance}, $Z=R_\mathrm{L}+2\pi fL$. The value $R_\mathrm{stray}=0.45$\,\si{m\ohm} was significantly lower than the 1-2.5\,\si{m\ohm} measured by us using alternative configurations. Low stray resistance is crucial to achieving a near-ideal TES voltage bias. A total inductance of $L=(L_\mathrm{in}+L_\mathrm{stray})=108$ nH was measured. Using the manufacturer's value for the SQUID input impedance, $L_\mathrm{in} = 80$ nH, implies a stray inductance $L_\mathrm{stray}=28$\,nH. Knowledge of the total inductance is required for calculating the TES response times, as described in Section \ref{sec:ElectrothermalMod}.

\subsection{TES characterisation: thermal conductance and optical absorption}

\begin{figure}
\centering
\includegraphics[]{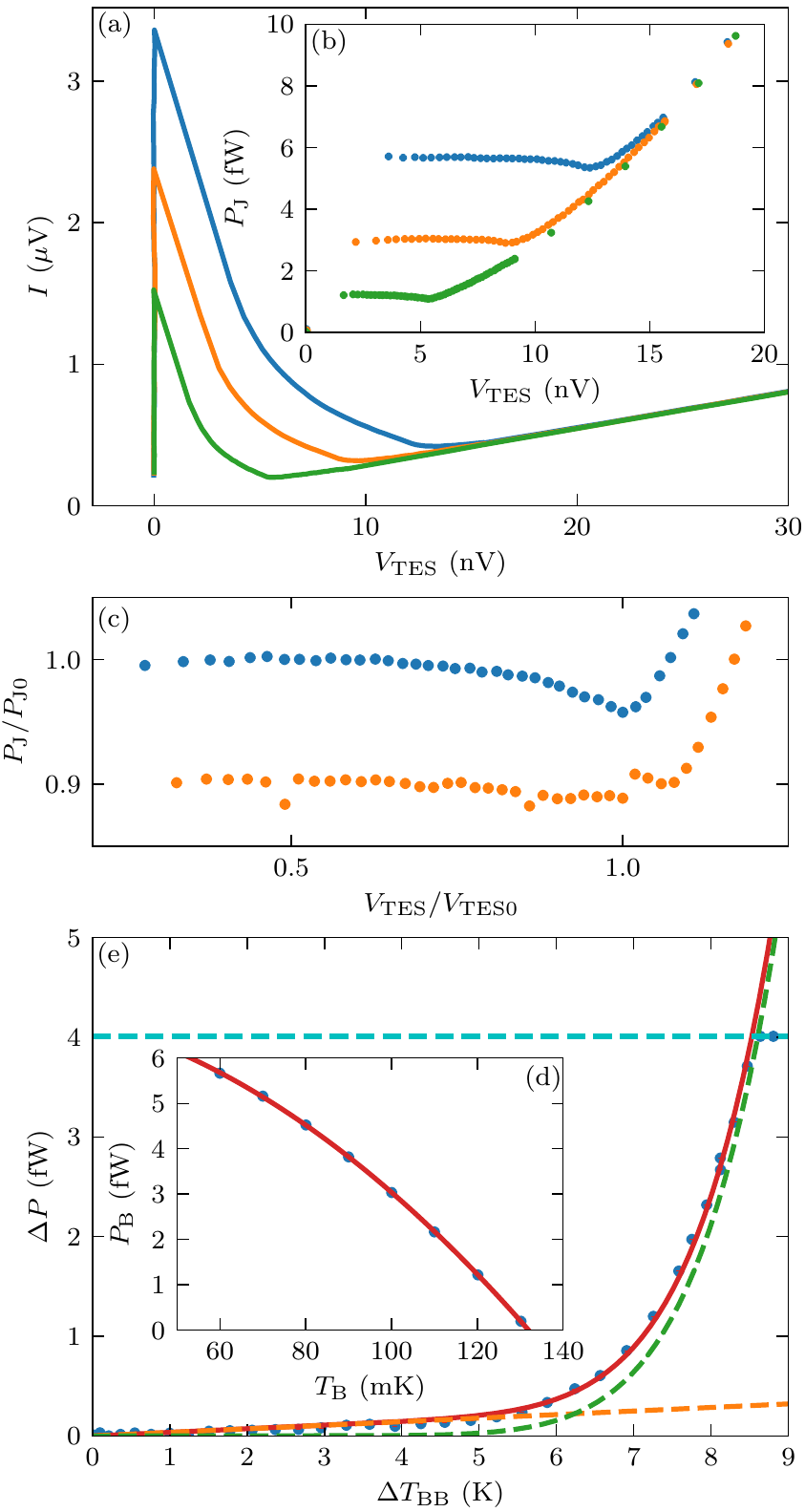}
\caption{\label{fig:GOpt} (a) TES current, $I$, against voltage across the TES, $V_\mathrm{TES}$ for a subset of bath temperatures, $T_\mathrm{B}$: 60 mK (blue), 100 mK (orange) and 120 mK (green). (b) Corresponding Joule power $P_\mathrm{J}$ against $V_\mathrm{TES}$ for these bath temperatures. (c) Net thermal power to the heat bath, $P_\mathrm{B}$, equal to $P_\mathrm{J}$ averaged over the transition, against $T_\mathrm{B}$. (d) $P_\mathrm{J}$ normalised to its mean value over the transition $P_\mathrm{J0}$ against $V_\mathrm{TES}$ normalised to its value at the normal end of the transition, for the tested M-band TES in the test facility (blue) and a longer wavelength device in a previous module design (orange). Data points for the longer wavelength device have been displaced to $P_\mathrm{J}/P_\mathrm{J0}=0.9$ for clarity. (e) Absorbed optical power $\Delta P$ against change in hot load temperature, $\Delta T_\mathrm{BB}$. Saturation power at $T_\mathrm{B}=90$ mK is indicated in cyan and model optical power $\Delta P_\mathrm{mod} = \alpha T_\mathrm{BB} + \eta_\mathrm{opt}\Delta P_\mathrm{th}$ in red, where $\alpha$ is a linear stray light coefficient, $\eta_\mathrm{opt}$ is the TES optical efficiency and $\Delta P_\mathrm{th}$ is the theoretical total throughput. Contributions $\alpha T_\mathrm{BB}$ and $\eta_\mathrm{opt}\Delta P_\mathrm{th}$ are shown in orange, dashed, and green, dashed, respectively.}
\end{figure}

Figure \ref{fig:GOpt}(a) shows the TES $I-V_\mathrm{TES}$ curve, for a selection of bath temperatures $T_\mathrm{B}$, where $T_\mathrm{B}$ is taken to be the temperature of the detector housing controlled through the residual current in the ADR magnet. The corresponding dissipated Joule power, $P_\mathrm{J}=IV_\mathrm{TES}$, is shown in Fig. \ref{fig:GOpt}(b), from $T_\mathrm{B}=60$ mK (blue, top) to $T_\mathrm{B}=120$ mK (green, bottom).

Figure \ref{fig:GOpt}(c) compares a representative $P_\mathrm{J}-V_\mathrm{TES}$ plateau measured using the optical test facility with a previous measurement on a different TES, in a previous test system without custom magnetic shielding. Power has been normalised to its mean value across the transition and voltage to its value at the start of the transition to allow a comparison between the two TESs, which were designed with different thermal conductances and transition temperatures. The earlier measurement has been displaced on the power axis for clarity in Fig. \ref{fig:GOpt}(c). The measurement from the earlier test system exhibits significant transient current drops, caused by low-frequency electromagnetic interference. These artefacts are completely absent in the new measurements despite the five times greater sensitivity of the TES, suggesting that the magnetic shielding and anchored wiring of the test facility were effective in eliminating pick up.

In the steady state, the net power flow from the TES island to the heat bath, $P_\mathrm{B}$, is equal to the Joule power dissipated in the bilayer, $P_\mathrm{J}$, which is approximately constant across the transition as evident in Fig. \ref{fig:GOpt}(b). Figure \ref{fig:GOpt}(d) shows $P_\mathrm{B}$ calculated as mean $P_\mathrm{J}$ over the transition for each $T_\mathrm{B}$.  The power flow to the bath is described by the expression
\begin{equation}
P_\mathrm{B} = K(T_\mathrm{C}^n-T_\mathrm{B}^n),
\label{eq:PKn}
\end{equation}
where $K$ is a parameter that scales the overall heat flux, and $n\sim 2-4$ is a factor that reflects the fundamental phonon-transport mechanism in the legs. Fitting Eq. \ref{eq:PKn} to the data shown in Fig. \ref{fig:GOpt}(d) results in values $n=1.98$, $K=400$ $\mathrm{fW/K^{n}}$ and $T_\mathrm{C}=131.9$ mK. The thermal conductance $G$ to the heat bath is then given by $G=nKT_\mathrm{C}^{(n-1)}=108.0$ fW/K, from which the noise equivalent power (NEP) may be calculated as $\mathrm{NEP}=\sqrt{4k_\mathrm{B}GT_\mathrm{C}^2}=0.32$ $\mathrm{aW/\sqrt{Hz}}$, where $k_\mathrm{B}$ is the Boltzmann constant. This exceptionally low NEP combined with the saturation power corresponding to $P_\mathrm{B}$ at a given temperature, for instance 3.8\,fW at 90\,mK, emphasise the utility of the device.

A crucial aspect of detector characterisation is determining the optical absorption efficiency, for which it is imperative that out-of-band stray radiation is minimised. As described in Section \ref{subsec:EnhancedModule}, the far-infrared absorbers were illuminated with a blackbody hot load of temperature $T_\mathrm{BB}$, through band-defining filters. As $T_\mathrm{BB}$ is increased above its base temperature $T_\mathrm{BB0}\approx3.3$K, more optical power is absorbed causing a reduction in Joule power dissipation. The absorbed power is given by
\begin{equation}
\Delta P = P_\mathrm{J}(T_\mathrm{BB0})-P_\mathrm{J}(T_\mathrm{BB}).
\end{equation}

Figure \ref{fig:GOpt}(e) shows $\Delta P$ against change in $T_\mathrm{BB}$, $\Delta T_\mathrm{BB}=T_\mathrm{BB}-T_\mathrm{BB0}$, with the bath temperature $T_\mathrm{B}$ maintained at 90 mK. When the absorbed power is equal to the Joule power at $T_\mathrm{BB0}$, the TES reaches its normal state. This is the differential saturation power of the TES relative to the incident power at $T_\mathrm{BB} = T_\mathrm{BB0}$, indicated on Fig. \ref{fig:GOpt}(e) (dashed cyan line and final added point). Whilst the rapid increase in $\Delta P$ above $\Delta T_\mathrm{BB}\approx5$ K is characteristic of M-band power absorption, there is evidence of a very slight additional power contribution at lower $T_\mathrm{BB}$, linear with $\Delta T_\mathrm{BB}$. This is attributed to long-wavelength stray light, as was observed to a far greater extent in our earlier test apparatus \cite{williams2018ultra}. The absorbed power $\Delta P$ may therefore by modelled as
\begin{equation}
\Delta P_\mathrm{mod} = \alpha\Delta T_\mathrm{BB}+\eta_\mathrm{opt}\Delta P_\mathrm{th},
\label{eq:DeltaPModGeo}
\end{equation}
where $\alpha$ is a stray light coefficient and $\eta_\mathrm{opt}$ is the detector optical efficiency. The theoretical overall optical throughput to the detectors, $P_\mathrm{th}(T_\mathrm{BB})$ may be calculated as

\begin{equation}
P_\mathrm{th} = A\Omega \int_{\lambda_\mathrm{min}}^{\lambda_\mathrm{max}} \eta_\mathrm{filters}\frac{2hc^2}{\lambda^{5}} \frac{1}{\mathrm{e}^\frac{hc}{\lambda k_\mathrm{B}T_\mathrm{BB}}-1} \, \mathrm{d}\lambda,
\label{eq:PthOpt}
\end{equation}
where $A$ is the TES absorber area, $\Omega$ is the solid angle subtended by the optical aperture, $\lambda_\mathrm{min}=50$ \si{\micro\metre} and $\lambda_\mathrm{max}=160$ \si{\micro\metre}, $\eta_\mathrm{filters}$ is the product of the thermal and band-pass filter transmission coeffcients, $h$ is Planck's constant, $k_\mathrm{B}$ is the Boltzmann constant and $c$ is the speed of light. Equation \ref{eq:DeltaPModGeo} was fitted to the measured data, giving $\Delta P_\mathrm{mod}$ shown in Fig. \ref{fig:GOpt}(e) (red) with $\alpha = 0.035$ and $\eta_\mathrm{opt}=1.08$. This near-unity optical efficiency indicates close to ideal optical performance of the TES absorber-backshort assembly. The slight overestimation may be due, for example, to some small systematic error in bias and readout circuit parameters, or in the dimensions featuring in Eq. \ref{eq:PthOpt}. The stray light coefficient $\alpha$ is approximately three times lower than that measured for this TES in our early test facility designs, where the hot load was thermally connected to a cylindrical radiation shield surrounding the 50 mK components \cite{williams2018ultra}. A significant reduction in stray light has therefore been achieved in this test facility compared to earlier configurations, especially given that the thermal path from the hot load must now pass within the closed shielding in close proximity to the detector enclosure before reaching the 3 K plate of the cryostat. The origin of the remaining contribution is the subject of further investigation, but nevertheless the quality of the data is pleasing given the exceedingly low powers involved.

\subsection{Calibration of optical modulator}

\begin{figure}
\centering
\includegraphics[]{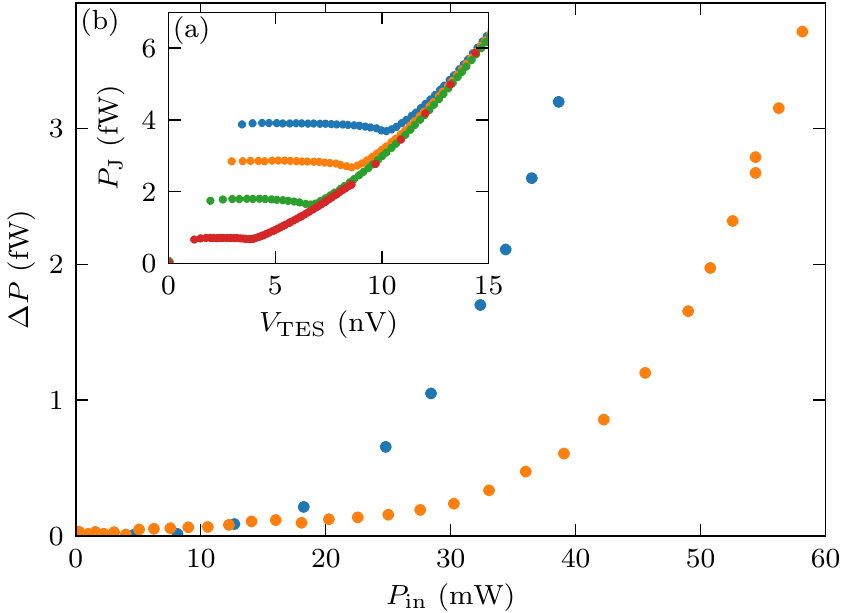}
\caption{\label{fig:CROpt} (a) Joule power $P_\mathrm{J}$ against voltage across the TES, $V_\mathrm{TES}$, for increasing power input to the chip resistor from 0 W (blue) to 38.6 mW (red). (b) Absorbed optical power, $\Delta P$, against input $P_\mathrm{in}$ to the chip resistor (blue) and to the hot load (orange).}
\end{figure}

To investigate the optical coupling between the resistive chip modulator and the detectors, the optical power absorbed by a TES was measured whilst the electrical power dissipated in the chip resistor was increased, with the primary hot load remaining at base temperature. Figure \ref{fig:CROpt}(a) shows $P_\mathrm{J}-V_\mathrm{TES}$ for a subset of electrical powers from $0-40$ mW, demonstrating progressive absorption across the full dynamic range up to TES saturation. Absorbed power $\Delta P$ is shown against electrical power $P_\mathrm{in}$ in Fig. \ref{fig:CROpt} for the modulator (blue) compared to that of the hot load (orange). The functional form of the increase in $\Delta P$ with $P_\mathrm{in}$ for the optical modulator reflects that of the hot load, indicating in-band illumination of the detectors, and showing sensitivity to $\approx$100 aW power from the modulator, even with a wide readout bandwidth.

\subsection{TES current response to modulation in bias voltage and optical power}
\label{subsec:Risetimes}

The test system allowed us to compare the response of a TES to a small step in bias voltage, $\delta V$, with that of an optical pulse, $\delta P$. The TES was biased within its transition, at 33$\%$ of its normal state resistance without optical loading, corresponding to a bias voltage $V\approx7.6$\,nV. To obtain the response to bias voltage modulation, a low amplitude square wave was superimposed on the bias input, giving $\delta V\approx0.13$\,nV. The change in TES current as a function of time, $\delta I(t)$, was averaged over 40 periods of the square wave, separately for the leading and trailing edges, corresponding to an increase in $V$ and a return to initial $V$ respectively.

\begin{figure*}
\centering
\includegraphics[]{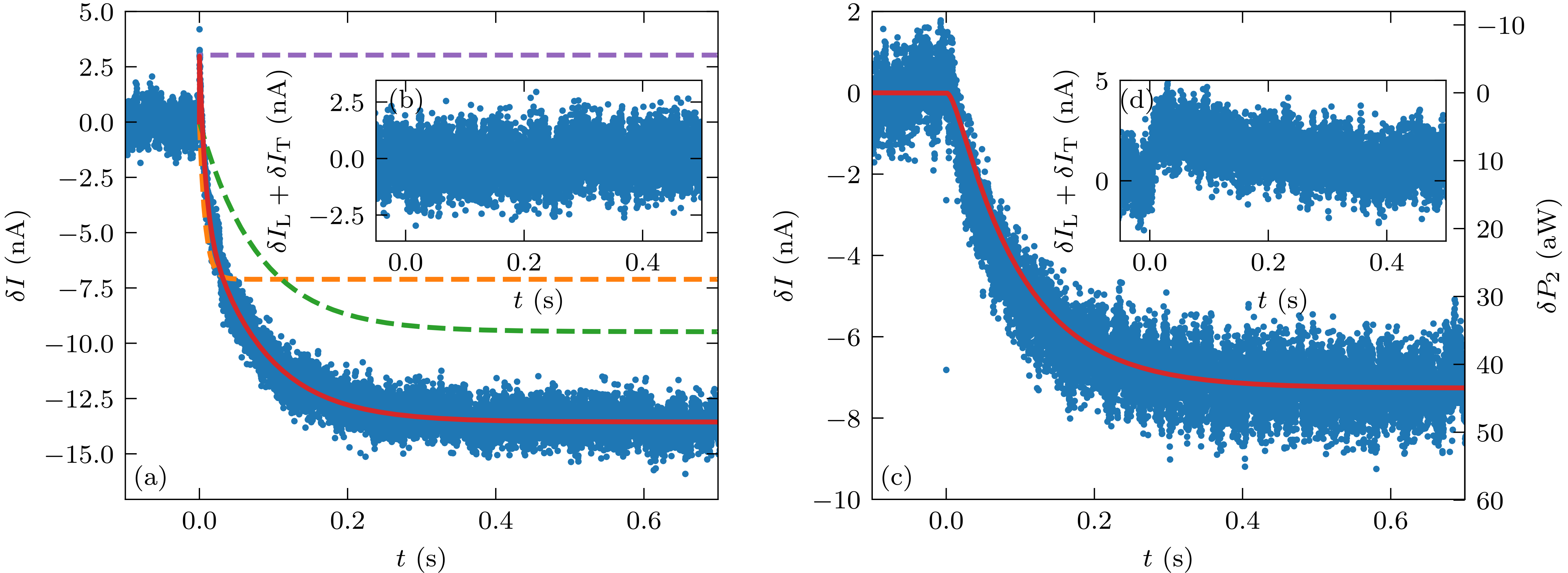}
\caption{\label{fig:RtElOpt} (a) Change in TES current, $\delta I$, with time, $t$, in response to a small increase in bias voltage, $\delta V$ at $t=0$, with the blackbody hot load at base temperature, $T_\mathrm{BB}=T_\mathrm{BB0}$. A decomposition into exponential contributions according to Eq. \ref{eq:RtSSMod1}, with fitted amplitudes and time constants, is shown with additive terms (lilac, orange, green) and sum (red). (b) Sum of $\delta I$ for leading and trailing $\delta V$, $\delta I_\mathrm{L}$ and $\delta I_\mathrm{T}$ respectively. (c) $\delta I(t)$ for a positive step in optical power input, $\delta P$, at $t=0$, for $T_\mathrm{BB}=T_\mathrm{BB0}$. Simulated $\delta I$ for power input $\delta P_2$ to the second heat capacity is shown (red), scaled to match measured final $\delta I$. Right-hand axis shows $\delta P_2$ calculated from this scaling factor. (d) $\delta I_\mathrm{L}+\delta I_\mathrm{T}$ for the leading and trailing optical responses.}
\end{figure*}

Figure \ref{fig:RtElOpt}(a) shows the change in TES current, $\delta I(t)$, on the leading edge of the voltage pulse, with the blackbody hot load at base temperature, $T_\mathrm{BB} = T_\mathrm{BB0}=3.3$\,K. The initial abrupt increase in current following the voltage step at $t=0$ is due to the electrical response of the TES and bias circuit. This is followed by a slow relaxation towards the new, lower, steady state current under electrothermal feedback, corresponding to a negative final current change $\delta I_\mathrm{f}$ for positive $\delta V$. The change in Joule power, $\delta P_\mathrm{J}$, corresponding to $\delta I_\mathrm{f}\approx13.8$\,nA is approximately 1.5\,aW. Therefore, although these plots look `noisy', each tick mark on the vertical current axis, corresponds to a power change of approximately 0.3\,aW. Taking this value as an estimate of the noise floor and using the saturation power of 4\,fW from earlier gives a dynamic range of 41\,dB in this case. The signal to noise ratio and dynamic range achieved are more than acceptable given the minute changes in TES steady state being considered.

To quantify the dependence of $\delta I(t)$ on absorbed optical power, each measured response may be decomposed into a sum of exponential contributions, to be compared with those predicted by small-signal electrothermal theory. A function having the form of Eq. \ref{eq:RtSSMod1} was therefore fitted to the measured data, but with the amplitudes, $a_\mathrm{i1}$ for $i=1,2,3$ and time constants, $\tau_\mathrm{i}$, as free parameters rather than being calculated through Eqs. \ref{eq:Matrix} and \ref{eq:bSS}. These parameters were therefore not constrained to any particular inter-relation or progression with $T_\mathrm{BB}$.

\begin{table*}
\caption{\label{tab:AmpTau} First elements $a_\mathrm{i1}$ of the amplitude vectors $\mathbf{a_\mathrm{i}}$, and time constants $\tau_\mathrm{i}$, where $i=1,2,3$, parameters of Equation \ref{eq:RtSSMod1} forming a functional representation of $\delta I$. M/S indicates values (M) acquired from free fitting to measured $\delta I$ at $T_\mathrm{BB}=T_\mathrm{BB0}$, and values (S) simulated from the small signal electrothermal model, with amplitudes given by Equations \ref{eq:RtSSMod2} and \ref{eq:bSS}, and time constants from Equation \ref{eq:Diff_Matrix}. Step type is indicated as being either in bias voltage, $\delta V$, or optical power, $\delta P$, with $\delta P_2$ indicating power input to the second heat capacity in simulations. Leading edge (L) describes positive $\delta V$ or $\delta P$, and trailing edge T negative $\delta V$ or $\delta P$ returning to initial bias voltage or optical power.}
\begin{ruledtabular}
\begin{tabular}{ccccccccc}
M/S & Step & L/T & $a_{11}$ (nA) & $a_{21}$ (nA) & $a_{31}$ (nA) & $\tau_1$ (\si{\micro s}) & $\tau_2$ (ms) & $\tau_3$ (ms) \\
\hline
M & $\delta V$ & L & 3.03 & -7.11 & -9.48 & 7 & 7.8 & 79.9 \\
M & $\delta V$ & T & -3.46 & 7.73 & 9.38 & 7 & 9.1 & 88.3 \\
S & $\delta V$ & L & 2.94 &- 8.09 &- 8.63 & 2.33 & 10.1 & 94.7 \\
M & $\delta P$ & L & - & - & -7.41 & - & - & 84.5 \\
M & $\delta P$ & T & - & 3.02 & 5.54 & - & 7.4 & 57.9\\
S & $\delta P_2$ & L & 0 & 0.86 & -8.13 & 2.33 & 10.1 & 94.7 \\
\end{tabular}
\end{ruledtabular}
\end{table*}

Figure \ref{fig:RtElOpt}(a) shows a least-squares fit to the measured response (solid, red), with additive components $(\mathrm{a_\mathrm{i1}}-\mathrm{a_\mathrm{i1}e}^{-t/\tau_\mathrm{i}})$ (lilac, orange, green). The shortest time constant, corresponding to the electrical contribution, was fixed at $\tau_1=7$\,\si{\micro s} $\approx L/(R_\mathrm{L}+R_0)$. The precise value assumed for $\tau_1$ has negligible effect on the fitting as it is 3-4 orders of magnitude smaller than the next largest time constant, and only affects a small number of points due to the sampling time. Fitted parameters $a_\mathrm{i1}$ and $\tau_\mathrm{i}$ are listed in Table \ref{tab:AmpTau}, indicating contributions of comparable amplitude from terms in two thermal constants $\tau_2$ and $\tau_3$, where $\tau_3\approx10\tau_2$, rather than one as would be expected from a single heat capacity model.

The heat capacities, $C_1$ and $C_2$, and linking thermal conductance $G_{12}$, were estimated by fitting the small signal electrothermal model described by Eqs. \ref{eq:RtSSMod1}-\ref{eq:bSS} to $\delta I(t)$ at $T_\mathrm{BB}=T_\mathrm{BB0}$. Steady state values $I_0$, $R_0$, $T_{10}$, $T_{20}$ were simulated based on a parametric $R(T_1,I)$ surface \cite{rostem2007multitone} to capture coarse large signal behaviour. The resistance-current sensitivity $\beta$ was derived from measurements of the circuit impedance in the high frequency limit \cite{lindeman2004impedance}, and the relationship $\alpha=100\beta$ assumed \cite{goldie2009thermal}. It was not the aim of this work to extract highly accurate electrothermal parameters or heat capacities for this particular device, but rather to interpret the forms of measured response profiles in the context of TES electrothermal behaviour.

The values $C_1=34.7\pm0.4$\,fJ/K, $C_2=70.5\pm0.4$\,fJ/K and $G_{12}=1.33\pm0.02$\,pW/K were obtained. That $C_1$, containing the bilayer, is a factor of two smaller than $C_2$ may offer some support to the notion that $C_2$ is associated with the larger volume of $\mathrm{SiN_x}$ underlying the absorber. A comparable ratio $C_2/C_1=1.6$ was previously estimated from the constituent material properties of the regions of the device island featuring the absorber and the TES. Corresponding amplitudes $a_\mathrm{i1}$ and time constants $\tau_\mathrm{i}$ are listed in Table \ref{tab:AmpTau}, and show strong similarity with those derived from the free amplitude and time constant fit shown in Fig. \ref{fig:RtElOpt}(a).


In the limit of small changes in $I$, $R$, $T_1$ and $T_2$, and heat capacities and thermal conductances independent of temperature, it is expected that $\delta I(t)$ for positive $\delta V$ is equal to the negative of that for negative $\delta V$. Figure \ref{fig:RtElOpt}(b) shows the sum of $\delta I$ for leading and trailing $\delta V$, indicating that the absolute responses are essentially identical and therefore validating the use of the small signal limit.

To measure the response $\delta I(t)$ for a small step in optical power $\delta P$, the optical modulator was used to provide a small modulation, superimposed on a steady state background from the hot load. A voltage square wave was applied to the modulator, and the TES current averaged over 40 leading edges, $\delta P$ positive, and 40 trailing edges, as the optical perturbation returns to zero. An unexpected contribution to $\delta I(t)$ was observed, with a time constant around 3 s, considerably longer than any expected either from the TES or optical modulator. It is suspected that this contribution originates from stray light: for example, from some element of the module warming slowly and radiating as electrical power is applied to the modulator chip resistor. The source of this stray light has not yet been identified; however, possible candidates include the outer hot load housing to which the chip resistor is anchored, or the optical filters themselves, which rely on their metallic patterning to heat sink the polypropylene substrate.

Figure \ref{fig:RtElOpt}(c) shows $\delta I(t)$ for positive $\delta P$, for $T_\mathrm{BB}=T_\mathrm{BB0}$, with the stray light contribution subtracted. The rapid drop in current over the first few hundred milliseconds is characteristic of the TES electrothermal response to a step in incident power, demonstrating that the optical modulator is capable of delivering a modulated signal with steps that are abrupt on the timescale of the TES response.

Table \ref{tab:AmpTau} lists freely varying amplitudes and time constants obtained through fitting Eq. \ref{eq:RtSSMod1} to $\delta I(t)$. In the case of the leading edge optical data, a convincing fit was obtained with a single thermal time constant $\tau_3$. This behaviour is consistent with power input to the second heat capacity, $\delta P_2$, for which the longest time constant dominates, since thermal flux must traverse the weak thermal link $G_{12}$ before a response can be seen in the TES. The modelled response for power $\delta P_2$ into $C_2$ is shown in Fig. \ref{fig:RtElOpt}(c) scaled to match measured $\delta I_\mathrm{f}$ and using $C_1$, $C_2$ and $G_{12}$ derived from the voltage step measurement at $T_\mathrm{BB0}$ without further fitting. The value of $\delta P_2$ derived from the global scaling factor is 43.5\,aW, half the 87.4\,aW predicted from the $I-V_\mathrm{TES}$ measurement of Fig. \ref{fig:CROpt} for the same input power to the optical modulator. However, since the fitted stray light amplitude, 8.93\,nA, is almost equal to the amplitude $|a_{31}|=7.41$ nA, it is likely that M-band power accounts for roughly 50$\%$ of the 87.4\,aW total, with the remainder arising from stray light absorption at long timescales, rendering $\delta P_2=43.5$\,aW a reasonable M-band power amplitude.

Absorbed optical power is shown on the additional axis to the right of Fig. \ref{fig:RtElOpt}(c), assuming $\delta P_2=43.5$\,aW and linearity of $\delta I$ with $\delta P$ in the small signal limit. This emphasises that not only is the TES eminently capable of detecting the extremely small signal applied in the measurement shown, but would register absorbed power down to around 10 aW whilst remaining above the noise associated with the present experimental system.

That the modelled response for power applied to the second heat capacitance is able to closely reproduce the measured $\delta I(t)$ suggests that the superconducting bilayer is indeed weakly thermally coupled to the site of optical absorption in this device, resulting in a slower current response to a change in optical signal than to a change in bias voltage. Since optical absorption by the impedance-matched absorber is expected to dominate over direct absorption by the bilayer, this also supports the suggestion that $C_2$ represents the absorber within this simple model.

Contrary to expectation, a faster response was observed for optical $\delta I(t)$ on the trailing edge than the leading edge. This is reflected in the sum of the responses shown in Fig. \ref{fig:RtElOpt}(d), and in  the amplitudes $a_\mathrm{i1}$ and time constants $\tau_\mathrm{i}$ listed in Table \ref{tab:AmpTau}. In this case, $\delta I(t)$ is well described by two thermal time constants of comparable amplitude, in greater similarity to the electrical response than the leading edge optical. This asymmetry, as yet unexplained, would not have been revealed through measurements of TES response to bias voltage modulation alone.

\begin{figure*}
\centering
\includegraphics[]{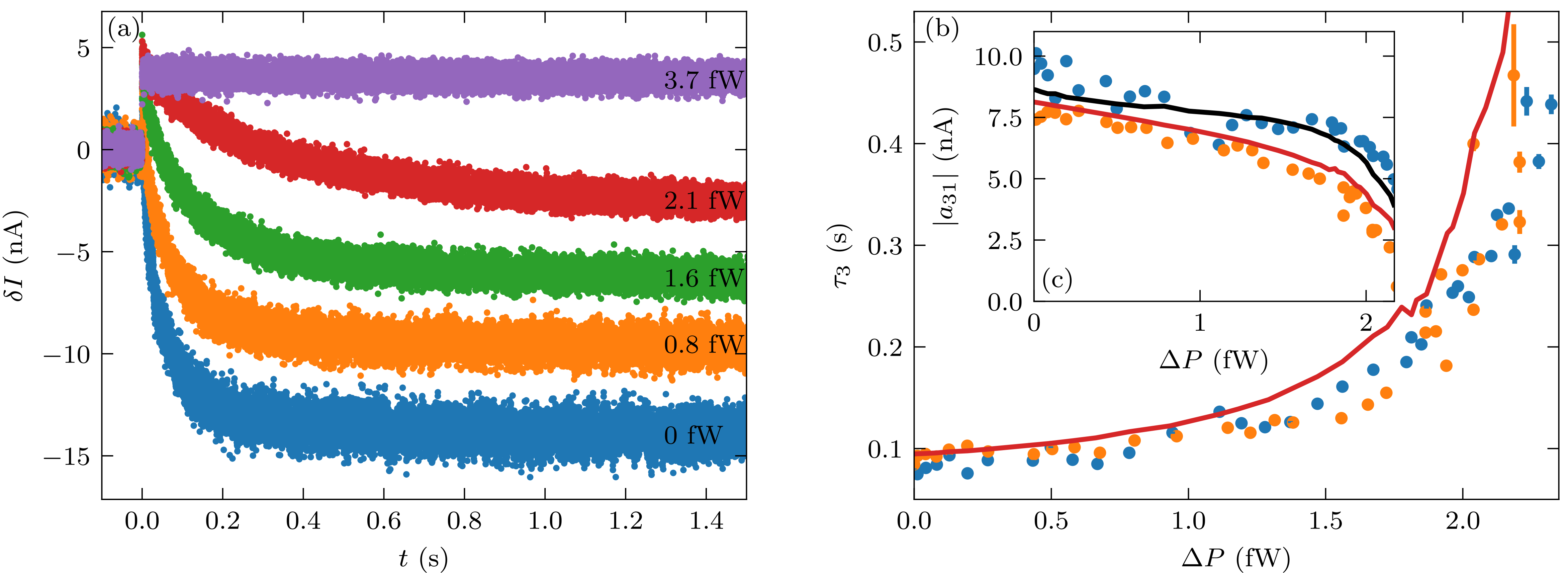}
\caption{\label{fig:RtElOptTBB} (a) Change in TES current, $\delta I$, with time, $t$, in response to a small increase in bias voltage, $\delta V$ at $t=0$, for absorbed background optical power $P$ from 0 fW (blue) to 3.7 fW (lilac), as indicated. (b) Fitted values for the longest time constant, $\tau_3$, for $\delta V$ (blue) and $\delta P$ (orange) against $P$, with simulated progression (red). (c) Absolute values of amplitudes $a_{31}$ associated with $\tau_3$ for $\delta V$ (blue) and $\delta P$ (orange), with simulated trends (red, black).}
\end{figure*}

A key functionality of the test facility is the ability to measure both the electrical and optical TES responses under background optical illumination from the blackbody hot load. Figure \ref{fig:RtElOptTBB}(a) shows the leading edge electrical response to a small step in bias voltage, as in Fig. \ref{fig:RtElOpt}(a), for a selection of blackbody hot load temperatures $T_\mathrm{BB}$ increasing from base temperature, $T_\mathrm{BB0}\approx3.3$ K (blue) to $T_\mathrm{BB}\approx12$ K (lilac). As $T_\mathrm{BB}$ increases, the magnitude of the final current change, $\delta I_\mathrm{f}$, decreases as TES resistance increases and approaches the normal state. Finally, at the highest $T_\mathrm{BB}$, the absorbed optical power, $\Delta P = 3.7$ fW as calculated from Eq. \ref{eq:DeltaPModGeo}, exceeds the saturation power of the device at this bias point and bath temperature $T_\mathrm{B}=90$ mK, and the TES returns to its normal state. In the absence of electrothermal feedback, the electrical time constant reverts to $L/(R_\mathrm{L}+R_n)$ and the TES behaves as a normal metal resistor with positive $\delta I_\mathrm{f}$.

It is evident from Fig. \ref{fig:RtElOptTBB}(a) that the time taken for the TES current to reach its new equilibrium increases with increasing optical power loading, as the resistance-temperature sensitivity $\alpha$ decreases, up until saturation and the vanishing of the electrothermal contributions to the response time. This effect is likewise observed in corresponding optical response data for increasing $T_\mathrm{BB}$.

Figure \ref{fig:RtElOptTBB}(b) shows freely fitted $\tau_3$, the longest electrothermal time constant, against $P$ for leading edge $\delta V$ and $\delta P$. As expected, good agreement is observed between $\tau_3$ for the electrical and optical responses. It is also possible to calculate the expected value of $\tau_3$ with absorbed power $P$ without further fitting, using the small signal thermal model, measured $\beta$ with $T_\mathrm{BB}$, and the values obtained for $C_1$, $C_2$ and $G_{12}$. This is also shown in Fig. \ref{fig:RtElOptTBB}(b), where the small kinks arise from random error in $\beta$. Measured $\tau_3$ is seen to approximately follow the simulated trend, demonstrating that the observed slowing of the response with optical power is indeed an expected consequence of TES electrothermal behaviour. By TES saturation, $\tau_3$ has increased to five times its initial value, considerably slowing $\delta I(t)$. This corresponds to an increase in the time taken for the current to relax to within 10$\%$ of its final steady state value of approximately seven times for $\delta V$ and five times for $\delta P$. This aspect of TES behaviour has important implications for the design of TES bolometers for optical instruments such as SAFARI, dictating, for example, optical power as a proportion of available saturation power that may be absorbed before the response time exceeds the lower limit set by the scanning of the Martin-Puplett interferometer. Figure \ref{fig:RtElOptTBB}(c) shows the fitted and simulated magnitude of the amplitude, $a_{31}$, associated with $\tau_3$, which decreases with increasing absorbed power as expected.

\section{Conclusions}

We have devised and implemented a cryogenic test facility for ultra-low-noise far-infrared transition edge sensors. These sensors are being developed for the SAFARI grating spectrometer on the cooled-aperture space telescope SPICA. Although the experimental arrangement is suitable for the whole of the SAFARI wavelength range, 34-230\,\si{\micro\meter}, we have focused on a representative set of measurements at 60-110\,\si{\micro\meter}.

A key feature of the optical configuration is its ability to measure optical efficiencies with respect to a few-mode beam having modal characteristics identical to those of an ideal imaging telescope. Moreover, the addition of a fast infrared thermal source allows the direct measurement of the temporal response of TESs to tiny changes in optical power. We have shown that it is possible to measure transient optical response in the presence of steady-state background loading, all the way up to detector saturation.

A crucial consideration in the design was the minimisation of stray light, and the maximisation of magnetic and electrical shielding. In the context of the detector module, considerable care went into enhancing sensitivity and optical efficiency, achieving a high performance thermal and mechanical design, ensuring repeatability through good metrology, and eliminating stray light.

Overall, the test facility performed well. The detector module had a base temperature of $T_\mathrm{B}=59.5$ mK, but was operated at 90.0 mK through stabilised feedback using a residual current in the ADR magnet, leading to a RMS temperature stability of \SI{140}{\micro\kelvin}. Even at this elevated temperature, the devices tested had an exceptionally low NEP, 0.32 $\mathrm{aW/\sqrt{Hz}}$, making them suitable for ultra-low-noise space applications. The optical efficiency was measured to be 108\%, near-ideal with slight overestimation possibly due to calibration, with an optical saturation power of 4 fW. This indicates that the Au micromachined Si backshorts and meshed $\beta$-phase Ta absorber functioned according to design. This was pleasing as the thickness of the Ta film had been decreased by 53$\%$ to compensate for the increased sheet resistance caused by meshing, thereby obtaining a effective impedance closely matched to free space. Despite considerable effort to eliminate stray light, a long-wavelength $<$ 2 mm leakage was discovered at low illumination powers. The source, and mechanism by which the stray light entered the well-sealed TES enclosure were not found.

The transient electrical and optical responses of the TES were measured, induced by modulating the Joule power dissipated in the bilayer and by generating a modulated optical signal using the fast thermal source, respectively. Because of the extreme sensitivity of the detectors, we were able to observe the temporal forms of the TES response to step changes in an optical power of only 43 aW.  These measurements revealed important, and unexpected, insights into behaviour. For example, we observed clear evidence of a long-time constant, 3 s,  heating of some optical component within the field of view. The measured functional forms imply the existence of a weak thermal link between the TES bilayer and the site of optical absorption, despite our attempt to ensure fast thermal response by connecting the TES bilayer directly to a Au thermalising bar around the periphery of the absorber. As predicted by electrothermal modelling, we found that the electrical and optical responses were slowed to several times their dark values when background optical loading was applied, up to TES saturation. These observations demonstrate the importance of being able to measure optical and electrical transient response when refining TES design.

Considerable care is needed to eliminate stray light, and indeed to avoid the scenario where a calibration load heats a nearby surface, which subsequently re-radiates over a relatively long time period. In the context of SAFARI, the associated impact on TES responsivity and response speed, intimately related to other time constants and scan speeds in the instrument, would be highly detrimental to performance. Stray light control, and excellent thermal design, both of the detector chips and instrument, are central to the operation of ultra-low-noise astronomy instruments.

\begin{acknowledgments}

The authors are grateful to the European Space Agency CTP programme, 4000107657/13/NL/HB, and the UK Space Agency NSTP programme, for funding this work. We would like to thank our colleagues at the European Space Agency, in particular Astrid Heske, Peter Verhoeve, and Kate Isaak, for their continued support. Magnetic modelling of the shielding configuration was carried out under the NSTP programme at the Earth Observation Navigation and Science Group of Airbus Defence and Space, and we would like to thank Christian Trenkel and Maike Lieser for their work throughout this project. Emily Williams is grateful for a PhD studentship from the NanoDTC, Cambridge, EP/L015978/1. 

The data that support the findings of this study are available from the corresponding author upon reasonable request.

\end{acknowledgments}

\bibliography{References}

\begin{thebibliography}{23}%
\makeatletter
\providecommand \@ifxundefined [1]{%
 \@ifx{#1\undefined}
}%
\providecommand \@ifnum [1]{%
 \ifnum #1\expandafter \@firstoftwo
 \else \expandafter \@secondoftwo
 \fi
}%
\providecommand \@ifx [1]{%
 \ifx #1\expandafter \@firstoftwo
 \else \expandafter \@secondoftwo
 \fi
}%
\providecommand \natexlab [1]{#1}%
\providecommand \enquote  [1]{``#1''}%
\providecommand \bibnamefont  [1]{#1}%
\providecommand \bibfnamefont [1]{#1}%
\providecommand \citenamefont [1]{#1}%
\providecommand \href@noop [0]{\@secondoftwo}%
\providecommand \href [0]{\begingroup \@sanitize@url \@href}%
\providecommand \@href[1]{\@@startlink{#1}\@@href}%
\providecommand \@@href[1]{\endgroup#1\@@endlink}%
\providecommand \@sanitize@url [0]{\catcode `\\12\catcode `\$12\catcode
  `\&12\catcode `\#12\catcode `\^12\catcode `\_12\catcode `\%12\relax}%
\providecommand \@@startlink[1]{}%
\providecommand \@@endlink[0]{}%
\providecommand \url  [0]{\begingroup\@sanitize@url \@url }%
\providecommand \@url [1]{\endgroup\@href {#1}{\urlprefix }}%
\providecommand \urlprefix  [0]{URL }%
\providecommand \Eprint [0]{\href }%
\providecommand \doibase [0]{http://dx.doi.org/}%
\providecommand \selectlanguage [0]{\@gobble}%
\providecommand \bibinfo  [0]{\@secondoftwo}%
\providecommand \bibfield  [0]{\@secondoftwo}%
\providecommand \translation [1]{[#1]}%
\providecommand \BibitemOpen [0]{}%
\providecommand \bibitemStop [0]{}%
\providecommand \bibitemNoStop [0]{.\EOS\space}%
\providecommand \EOS [0]{\spacefactor3000\relax}%
\providecommand \BibitemShut  [1]{\csname bibitem#1\endcsname}%
\let\auto@bib@innerbib\@empty
\bibitem [{\citenamefont {Irwin}\ and\ \citenamefont
  {Hilton}(2005)}]{IrwinChapter}%
  \BibitemOpen
  \bibfield  {author} {\bibinfo {author} {\bibfnamefont {K.~D.}\ \bibnamefont
  {Irwin}}\ and\ \bibinfo {author} {\bibfnamefont {G.~C.}\ \bibnamefont
  {Hilton}},\ }in\ \href@noop {} {\emph {\bibinfo {booktitle} {Cryogenic
  particle detection}}}\ (\bibinfo  {publisher} {Springer},\ \bibinfo {year}
  {2005})\ pp.\ \bibinfo {pages} {63--150}\BibitemShut {NoStop}%
\bibitem [{\citenamefont {Matsumura}\ \emph {et~al.}(2014)\citenamefont
  {Matsumura}, \citenamefont {Akiba}, \citenamefont {Borrill}, \citenamefont
  {Chinone}, \citenamefont {Dobbs}, \citenamefont {Fuke}, \citenamefont
  {Ghribi}, \citenamefont {Hasegawa}, \citenamefont {Hattori}, \citenamefont
  {Hattori} \emph {et~al.}}]{matsumura2014mission}%
  \BibitemOpen
  \bibfield  {author} {\bibinfo {author} {\bibfnamefont {T.}~\bibnamefont
  {Matsumura}}, \bibinfo {author} {\bibfnamefont {Y.}~\bibnamefont {Akiba}},
  \bibinfo {author} {\bibfnamefont {J.}~\bibnamefont {Borrill}}, \bibinfo
  {author} {\bibfnamefont {Y.}~\bibnamefont {Chinone}}, \bibinfo {author}
  {\bibfnamefont {M.}~\bibnamefont {Dobbs}}, \bibinfo {author} {\bibfnamefont
  {H.}~\bibnamefont {Fuke}}, \bibinfo {author} {\bibfnamefont {A.}~\bibnamefont
  {Ghribi}}, \bibinfo {author} {\bibfnamefont {M.}~\bibnamefont {Hasegawa}},
  \bibinfo {author} {\bibfnamefont {K.}~\bibnamefont {Hattori}}, \bibinfo
  {author} {\bibfnamefont {M.}~\bibnamefont {Hattori}},  \emph {et~al.},\
  }\href@noop {} {\bibfield  {journal} {\bibinfo  {journal} {J. Low Temp.
  Phys.}\ }\textbf {\bibinfo {volume} {176}},\ \bibinfo {pages} {733} (\bibinfo
  {year} {2014})}\BibitemShut {NoStop}%
\bibitem [{\citenamefont {Suzuki}\ \emph {et~al.}(2018)\citenamefont {Suzuki},
  \citenamefont {Ade}, \citenamefont {Akiba}, \citenamefont {Alonso},
  \citenamefont {Arnold}, \citenamefont {Aumont}, \citenamefont {Baccigalupi},
  \citenamefont {Barron}, \citenamefont {Basak}, \citenamefont {Beckman} \emph
  {et~al.}}]{suzuki2018litebird}%
  \BibitemOpen
  \bibfield  {author} {\bibinfo {author} {\bibfnamefont {A.}~\bibnamefont
  {Suzuki}}, \bibinfo {author} {\bibfnamefont {P.}~\bibnamefont {Ade}},
  \bibinfo {author} {\bibfnamefont {Y.}~\bibnamefont {Akiba}}, \bibinfo
  {author} {\bibfnamefont {D.}~\bibnamefont {Alonso}}, \bibinfo {author}
  {\bibfnamefont {K.}~\bibnamefont {Arnold}}, \bibinfo {author} {\bibfnamefont
  {J.}~\bibnamefont {Aumont}}, \bibinfo {author} {\bibfnamefont
  {C.}~\bibnamefont {Baccigalupi}}, \bibinfo {author} {\bibfnamefont
  {D.}~\bibnamefont {Barron}}, \bibinfo {author} {\bibfnamefont
  {S.}~\bibnamefont {Basak}}, \bibinfo {author} {\bibfnamefont
  {S.}~\bibnamefont {Beckman}},  \emph {et~al.},\ }\href@noop {} {\bibfield
  {journal} {\bibinfo  {journal} {J. Low Temp. Phys.}\ }\textbf {\bibinfo
  {volume} {193}},\ \bibinfo {pages} {1048} (\bibinfo {year}
  {2018})}\BibitemShut {NoStop}%
\bibitem [{\citenamefont {Roelfsema}\ \emph {et~al.}(2018)\citenamefont
  {Roelfsema}, \citenamefont {Shibai}, \citenamefont {Armus}, \citenamefont
  {Arrazola}, \citenamefont {Audard}, \citenamefont {Audley}, \citenamefont
  {Bradford}, \citenamefont {Charles}, \citenamefont {Dieleman}, \citenamefont
  {Doi} \emph {et~al.}}]{roelfsema2018spica}%
  \BibitemOpen
  \bibfield  {author} {\bibinfo {author} {\bibfnamefont {P.}~\bibnamefont
  {Roelfsema}}, \bibinfo {author} {\bibfnamefont {H.}~\bibnamefont {Shibai}},
  \bibinfo {author} {\bibfnamefont {L.}~\bibnamefont {Armus}}, \bibinfo
  {author} {\bibfnamefont {D.}~\bibnamefont {Arrazola}}, \bibinfo {author}
  {\bibfnamefont {M.}~\bibnamefont {Audard}}, \bibinfo {author} {\bibfnamefont
  {M.}~\bibnamefont {Audley}}, \bibinfo {author} {\bibfnamefont
  {C.}~\bibnamefont {Bradford}}, \bibinfo {author} {\bibfnamefont
  {I.}~\bibnamefont {Charles}}, \bibinfo {author} {\bibfnamefont
  {P.}~\bibnamefont {Dieleman}}, \bibinfo {author} {\bibfnamefont
  {Y.}~\bibnamefont {Doi}},  \emph {et~al.},\ }\href@noop {} {\bibfield
  {journal} {\bibinfo  {journal} {Publications of the Astronomical Society of
  Australia}\ }\textbf {\bibinfo {volume} {35}} (\bibinfo {year}
  {2018})}\BibitemShut {NoStop}%
\bibitem [{\citenamefont {Audley}\ \emph {et~al.}(2018)\citenamefont {Audley},
  \citenamefont {de~Lange}, \citenamefont {Gao}, \citenamefont {Jackson},
  \citenamefont {Hijmering}, \citenamefont {Ridder}, \citenamefont {Bruijn},
  \citenamefont {Roelfsema}, \citenamefont {Ade}, \citenamefont {Withington}
  \emph {et~al.}}]{audley2018safari}%
  \BibitemOpen
  \bibfield  {author} {\bibinfo {author} {\bibfnamefont {M.~D.}\ \bibnamefont
  {Audley}}, \bibinfo {author} {\bibfnamefont {G.}~\bibnamefont {de~Lange}},
  \bibinfo {author} {\bibfnamefont {J.-R.}\ \bibnamefont {Gao}}, \bibinfo
  {author} {\bibfnamefont {B.~D.}\ \bibnamefont {Jackson}}, \bibinfo {author}
  {\bibfnamefont {R.~A.}\ \bibnamefont {Hijmering}}, \bibinfo {author}
  {\bibfnamefont {M.~L.}\ \bibnamefont {Ridder}}, \bibinfo {author}
  {\bibfnamefont {M.~P.}\ \bibnamefont {Bruijn}}, \bibinfo {author}
  {\bibfnamefont {P.~R.}\ \bibnamefont {Roelfsema}}, \bibinfo {author}
  {\bibfnamefont {P.~A.}\ \bibnamefont {Ade}}, \bibinfo {author} {\bibfnamefont
  {S.}~\bibnamefont {Withington}},  \emph {et~al.},\ }in\ \href@noop {} {\emph
  {\bibinfo {booktitle} {Millimeter, Submillimeter, and Far-Infrared Detectors
  and Instrumentation for Astronomy IX}}},\ Vol.\ \bibinfo {volume} {10708}\
  (\bibinfo {organization} {Proc. SPIE},\ \bibinfo {year} {2018})\ p.\ \bibinfo
  {pages} {107080K}\BibitemShut {NoStop}%
\bibitem [{\citenamefont {Goldie}\ \emph {et~al.}(2016)\citenamefont {Goldie},
  \citenamefont {Glowacka}, \citenamefont {Withington}, \citenamefont {Chen},
  \citenamefont {Ade}, \citenamefont {Morozov}, \citenamefont {Sudiwala},
  \citenamefont {Trappe},\ and\ \citenamefont
  {Quaranta}}]{goldie2016performance}%
  \BibitemOpen
  \bibfield  {author} {\bibinfo {author} {\bibfnamefont {D.~J.}\ \bibnamefont
  {Goldie}}, \bibinfo {author} {\bibfnamefont {D.~M.}\ \bibnamefont
  {Glowacka}}, \bibinfo {author} {\bibfnamefont {S.}~\bibnamefont
  {Withington}}, \bibinfo {author} {\bibfnamefont {J.}~\bibnamefont {Chen}},
  \bibinfo {author} {\bibfnamefont {P.}~\bibnamefont {Ade}}, \bibinfo {author}
  {\bibfnamefont {D.}~\bibnamefont {Morozov}}, \bibinfo {author} {\bibfnamefont
  {R.}~\bibnamefont {Sudiwala}}, \bibinfo {author} {\bibfnamefont
  {N.}~\bibnamefont {Trappe}}, \ and\ \bibinfo {author} {\bibfnamefont
  {O.}~\bibnamefont {Quaranta}},\ }in\ \href@noop {} {\emph {\bibinfo
  {booktitle} {Millimeter, Submillimeter, and Far-Infrared Detectors and
  Instrumentation for Astronomy VIII}}},\ Vol.\ \bibinfo {volume} {9914}\
  (\bibinfo {organization} {Proc. SPIE},\ \bibinfo {year} {2016})\ p.\ \bibinfo
  {pages} {99140A}\BibitemShut {NoStop}%
\bibitem [{\citenamefont {Audley}\ \emph {et~al.}(2014)\citenamefont {Audley},
  \citenamefont {De~Lange}, \citenamefont {Ranjan}, \citenamefont {Gao},
  \citenamefont {Khosropanah}, \citenamefont {Ridder}, \citenamefont
  {Mauskopf}, \citenamefont {Morozov}, \citenamefont {Doherty}, \citenamefont
  {Trappe} \emph {et~al.}}]{audley2014measurements}%
  \BibitemOpen
  \bibfield  {author} {\bibinfo {author} {\bibfnamefont {M.}~\bibnamefont
  {Audley}}, \bibinfo {author} {\bibfnamefont {G.}~\bibnamefont {De~Lange}},
  \bibinfo {author} {\bibfnamefont {M.}~\bibnamefont {Ranjan}}, \bibinfo
  {author} {\bibfnamefont {J.-R.}\ \bibnamefont {Gao}}, \bibinfo {author}
  {\bibfnamefont {P.}~\bibnamefont {Khosropanah}}, \bibinfo {author}
  {\bibfnamefont {M.}~\bibnamefont {Ridder}}, \bibinfo {author} {\bibfnamefont
  {P.~D.}\ \bibnamefont {Mauskopf}}, \bibinfo {author} {\bibfnamefont
  {D.}~\bibnamefont {Morozov}}, \bibinfo {author} {\bibfnamefont
  {S.}~\bibnamefont {Doherty}}, \bibinfo {author} {\bibfnamefont
  {N.}~\bibnamefont {Trappe}},  \emph {et~al.},\ }\href@noop {} {\bibfield
  {journal} {\bibinfo  {journal} {J. Low Temp. Phys.}\ }\textbf {\bibinfo
  {volume} {176}},\ \bibinfo {pages} {755} (\bibinfo {year}
  {2014})}\BibitemShut {NoStop}%
\bibitem [{\citenamefont {Khosropanah}\ \emph {et~al.}(2014)\citenamefont
  {Khosropanah}, \citenamefont {Suzuki}, \citenamefont {Hijmering},
  \citenamefont {Ridder}, \citenamefont {Lindeman}, \citenamefont {Gao},\ and\
  \citenamefont {Hoevers}}]{khosropanah2014characterization}%
  \BibitemOpen
  \bibfield  {author} {\bibinfo {author} {\bibfnamefont {P.}~\bibnamefont
  {Khosropanah}}, \bibinfo {author} {\bibfnamefont {T.}~\bibnamefont {Suzuki}},
  \bibinfo {author} {\bibfnamefont {R.}~\bibnamefont {Hijmering}}, \bibinfo
  {author} {\bibfnamefont {M.}~\bibnamefont {Ridder}}, \bibinfo {author}
  {\bibfnamefont {M.}~\bibnamefont {Lindeman}}, \bibinfo {author}
  {\bibfnamefont {J.-R.}\ \bibnamefont {Gao}}, \ and\ \bibinfo {author}
  {\bibfnamefont {H.}~\bibnamefont {Hoevers}},\ }\href@noop {} {\bibfield
  {journal} {\bibinfo  {journal} {J. Low Temp. Phys.}\ }\textbf {\bibinfo
  {volume} {176}},\ \bibinfo {pages} {363} (\bibinfo {year}
  {2014})}\BibitemShut {NoStop}%
\bibitem [{\citenamefont {Barcons}\ \emph {et~al.}(2017)\citenamefont
  {Barcons}, \citenamefont {Barret}, \citenamefont {Decourchelle},
  \citenamefont {den Herder}, \citenamefont {Fabian}, \citenamefont
  {Matsumoto}, \citenamefont {Lumb}, \citenamefont {Nandra}, \citenamefont
  {Piro}, \citenamefont {Smith} \emph {et~al.}}]{barcons2017athena}%
  \BibitemOpen
  \bibfield  {author} {\bibinfo {author} {\bibfnamefont {X.}~\bibnamefont
  {Barcons}}, \bibinfo {author} {\bibfnamefont {D.}~\bibnamefont {Barret}},
  \bibinfo {author} {\bibfnamefont {A.}~\bibnamefont {Decourchelle}}, \bibinfo
  {author} {\bibfnamefont {J.}~\bibnamefont {den Herder}}, \bibinfo {author}
  {\bibfnamefont {A.}~\bibnamefont {Fabian}}, \bibinfo {author} {\bibfnamefont
  {H.}~\bibnamefont {Matsumoto}}, \bibinfo {author} {\bibfnamefont
  {D.}~\bibnamefont {Lumb}}, \bibinfo {author} {\bibfnamefont {K.}~\bibnamefont
  {Nandra}}, \bibinfo {author} {\bibfnamefont {L.}~\bibnamefont {Piro}},
  \bibinfo {author} {\bibfnamefont {R.}~\bibnamefont {Smith}},  \emph
  {et~al.},\ }\href@noop {} {\bibfield  {journal} {\bibinfo  {journal}
  {Astronomische Nachrichten}\ }\textbf {\bibinfo {volume} {338}},\ \bibinfo
  {pages} {153} (\bibinfo {year} {2017})}\BibitemShut {NoStop}%
\bibitem [{\citenamefont {Jackson}\ \emph {et~al.}(2016)\citenamefont
  {Jackson}, \citenamefont {Van~Weers}, \citenamefont {van~der Kuur},
  \citenamefont {den Hartog}, \citenamefont {Akamatsu}, \citenamefont {Argan},
  \citenamefont {Bandler}, \citenamefont {Barbera}, \citenamefont {Barret},
  \citenamefont {Bruijn} \emph {et~al.}}]{jackson2016focal}%
  \BibitemOpen
  \bibfield  {author} {\bibinfo {author} {\bibfnamefont {B.~D.}\ \bibnamefont
  {Jackson}}, \bibinfo {author} {\bibfnamefont {H.}~\bibnamefont {Van~Weers}},
  \bibinfo {author} {\bibfnamefont {J.}~\bibnamefont {van~der Kuur}}, \bibinfo
  {author} {\bibfnamefont {R.}~\bibnamefont {den Hartog}}, \bibinfo {author}
  {\bibfnamefont {H.}~\bibnamefont {Akamatsu}}, \bibinfo {author}
  {\bibfnamefont {A.}~\bibnamefont {Argan}}, \bibinfo {author} {\bibfnamefont
  {S.}~\bibnamefont {Bandler}}, \bibinfo {author} {\bibfnamefont
  {M.}~\bibnamefont {Barbera}}, \bibinfo {author} {\bibfnamefont
  {D.}~\bibnamefont {Barret}}, \bibinfo {author} {\bibfnamefont
  {M.}~\bibnamefont {Bruijn}},  \emph {et~al.},\ }in\ \href@noop {} {\emph
  {\bibinfo {booktitle} {Space Telescopes and Instrumentation 2016: Ultraviolet
  to Gamma Ray}}},\ Vol.\ \bibinfo {volume} {9905}\ (\bibinfo {organization}
  {Proc. SPIE},\ \bibinfo {year} {2016})\ p.\ \bibinfo {pages}
  {99052I}\BibitemShut {NoStop}%
\bibitem [{\citenamefont {Gottardi}\ \emph {et~al.}(2016)\citenamefont
  {Gottardi}, \citenamefont {Akamatsu}, \citenamefont {Bruijn}, \citenamefont
  {Den~Hartog}, \citenamefont {den Herder}, \citenamefont {Jackson},
  \citenamefont {Kiviranta}, \citenamefont {van~der Kuur},\ and\ \citenamefont
  {van Weers}}]{gottardi2016development}%
  \BibitemOpen
  \bibfield  {author} {\bibinfo {author} {\bibfnamefont {L.}~\bibnamefont
  {Gottardi}}, \bibinfo {author} {\bibfnamefont {H.}~\bibnamefont {Akamatsu}},
  \bibinfo {author} {\bibfnamefont {M.~P.}\ \bibnamefont {Bruijn}}, \bibinfo
  {author} {\bibfnamefont {R.}~\bibnamefont {Den~Hartog}}, \bibinfo {author}
  {\bibfnamefont {J.-W.}\ \bibnamefont {den Herder}}, \bibinfo {author}
  {\bibfnamefont {B.}~\bibnamefont {Jackson}}, \bibinfo {author} {\bibfnamefont
  {M.}~\bibnamefont {Kiviranta}}, \bibinfo {author} {\bibfnamefont
  {J.}~\bibnamefont {van~der Kuur}}, \ and\ \bibinfo {author} {\bibfnamefont
  {H.}~\bibnamefont {van Weers}},\ }\href@noop {} {\bibfield  {journal}
  {\bibinfo  {journal} {Nucl. Instrum. Methods Phys. Res}\ }\textbf {\bibinfo
  {volume} {824}},\ \bibinfo {pages} {622} (\bibinfo {year}
  {2016})}\BibitemShut {NoStop}%
\bibitem [{\citenamefont {Jackson}\ \emph {et~al.}(2011)\citenamefont
  {Jackson}, \citenamefont {De~Korte}, \citenamefont {Van~der Kuur},
  \citenamefont {Mauskopf}, \citenamefont {Beyer}, \citenamefont {Bruijn},
  \citenamefont {Cros}, \citenamefont {Gao}, \citenamefont {Griffin},
  \citenamefont {Den~Hartog} \emph {et~al.}}]{jackson2011spica}%
  \BibitemOpen
  \bibfield  {author} {\bibinfo {author} {\bibfnamefont {B.}~\bibnamefont
  {Jackson}}, \bibinfo {author} {\bibfnamefont {P.}~\bibnamefont {De~Korte}},
  \bibinfo {author} {\bibfnamefont {J.}~\bibnamefont {Van~der Kuur}}, \bibinfo
  {author} {\bibfnamefont {P.}~\bibnamefont {Mauskopf}}, \bibinfo {author}
  {\bibfnamefont {J.}~\bibnamefont {Beyer}}, \bibinfo {author} {\bibfnamefont
  {M.}~\bibnamefont {Bruijn}}, \bibinfo {author} {\bibfnamefont
  {A.}~\bibnamefont {Cros}}, \bibinfo {author} {\bibfnamefont {J.-R.}\
  \bibnamefont {Gao}}, \bibinfo {author} {\bibfnamefont {D.}~\bibnamefont
  {Griffin}}, \bibinfo {author} {\bibfnamefont {R.}~\bibnamefont {Den~Hartog}},
   \emph {et~al.},\ }\href@noop {} {\bibfield  {journal} {\bibinfo  {journal}
  {IEEE Trans. Terahertz Sci. Tech.}\ }\textbf {\bibinfo {volume} {2}},\
  \bibinfo {pages} {12} (\bibinfo {year} {2011})}\BibitemShut {NoStop}%
\bibitem [{\citenamefont {Den~Hartog}\ \emph {et~al.}(2012)\citenamefont
  {Den~Hartog}, \citenamefont {Audley}, \citenamefont {Beyer}, \citenamefont
  {Boersma}, \citenamefont {Bruijn}, \citenamefont {Gottardi}, \citenamefont
  {Hoevers}, \citenamefont {Hou}, \citenamefont {Keizer}, \citenamefont
  {Khosropanah} \emph {et~al.}}]{den2012low}%
  \BibitemOpen
  \bibfield  {author} {\bibinfo {author} {\bibfnamefont {R.}~\bibnamefont
  {Den~Hartog}}, \bibinfo {author} {\bibfnamefont {M.}~\bibnamefont {Audley}},
  \bibinfo {author} {\bibfnamefont {J.}~\bibnamefont {Beyer}}, \bibinfo
  {author} {\bibfnamefont {D.}~\bibnamefont {Boersma}}, \bibinfo {author}
  {\bibfnamefont {M.}~\bibnamefont {Bruijn}}, \bibinfo {author} {\bibfnamefont
  {L.}~\bibnamefont {Gottardi}}, \bibinfo {author} {\bibfnamefont
  {H.}~\bibnamefont {Hoevers}}, \bibinfo {author} {\bibfnamefont
  {R.}~\bibnamefont {Hou}}, \bibinfo {author} {\bibfnamefont {G.}~\bibnamefont
  {Keizer}}, \bibinfo {author} {\bibfnamefont {P.}~\bibnamefont {Khosropanah}},
   \emph {et~al.},\ }\href@noop {} {\bibfield  {journal} {\bibinfo  {journal}
  {J. Low Temp. Phys.}\ }\textbf {\bibinfo {volume} {167}},\ \bibinfo {pages}
  {652} (\bibinfo {year} {2012})}\BibitemShut {NoStop}%
\bibitem [{\citenamefont {van~der Kuur}\ \emph {et~al.}(2018)\citenamefont
  {van~der Kuur}, \citenamefont {Gottardi}, \citenamefont {Akamatsu},
  \citenamefont {Nieuwenhuizen}, \citenamefont {den Hartog},\ and\
  \citenamefont {Jackson}}]{van2018active}%
  \BibitemOpen
  \bibfield  {author} {\bibinfo {author} {\bibfnamefont {J.}~\bibnamefont
  {van~der Kuur}}, \bibinfo {author} {\bibfnamefont {L.}~\bibnamefont
  {Gottardi}}, \bibinfo {author} {\bibfnamefont {H.}~\bibnamefont {Akamatsu}},
  \bibinfo {author} {\bibfnamefont {A.}~\bibnamefont {Nieuwenhuizen}}, \bibinfo
  {author} {\bibfnamefont {R.}~\bibnamefont {den Hartog}}, \ and\ \bibinfo
  {author} {\bibfnamefont {B.}~\bibnamefont {Jackson}},\ }\href@noop {}
  {\bibfield  {journal} {\bibinfo  {journal} {J. Low Temp. Phys.}\ }\textbf
  {\bibinfo {volume} {193}},\ \bibinfo {pages} {626} (\bibinfo {year}
  {2018})}\BibitemShut {NoStop}%
\bibitem [{\citenamefont {Ade}\ \emph {et~al.}(2006)\citenamefont {Ade},
  \citenamefont {Pisano}, \citenamefont {Tucker},\ and\ \citenamefont
  {Weaver}}]{ade2006review}%
  \BibitemOpen
  \bibfield  {author} {\bibinfo {author} {\bibfnamefont {P.~A.}\ \bibnamefont
  {Ade}}, \bibinfo {author} {\bibfnamefont {G.}~\bibnamefont {Pisano}},
  \bibinfo {author} {\bibfnamefont {C.}~\bibnamefont {Tucker}}, \ and\ \bibinfo
  {author} {\bibfnamefont {S.}~\bibnamefont {Weaver}},\ }in\ \href@noop {}
  {\emph {\bibinfo {booktitle} {Millimeter and Submillimeter Detectors and
  Instrumentation for Astronomy III}}},\ Vol.\ \bibinfo {volume} {6275}\
  (\bibinfo {organization} {International Society for Optics and Photonics},\
  \bibinfo {year} {2006})\ p.\ \bibinfo {pages} {62750U}\BibitemShut {NoStop}%
\bibitem [{\citenamefont {Williams}\ \emph {et~al.}(2018)\citenamefont
  {Williams}, \citenamefont {Withington}, \citenamefont {Goldie}, \citenamefont
  {Thomas}, \citenamefont {Chen}, \citenamefont {Ade}, \citenamefont
  {Sudiwala}, \citenamefont {Walker},\ and\ \citenamefont
  {Trappe}}]{williams2018ultra}%
  \BibitemOpen
  \bibfield  {author} {\bibinfo {author} {\bibfnamefont {E.}~\bibnamefont
  {Williams}}, \bibinfo {author} {\bibfnamefont {S.}~\bibnamefont
  {Withington}}, \bibinfo {author} {\bibfnamefont {D.}~\bibnamefont {Goldie}},
  \bibinfo {author} {\bibfnamefont {C.}~\bibnamefont {Thomas}}, \bibinfo
  {author} {\bibfnamefont {J.}~\bibnamefont {Chen}}, \bibinfo {author}
  {\bibfnamefont {P.}~\bibnamefont {Ade}}, \bibinfo {author} {\bibfnamefont
  {R.}~\bibnamefont {Sudiwala}}, \bibinfo {author} {\bibfnamefont
  {I.}~\bibnamefont {Walker}}, \ and\ \bibinfo {author} {\bibfnamefont
  {N.}~\bibnamefont {Trappe}},\ }in\ \href@noop {} {\emph {\bibinfo {booktitle}
  {Millimeter, Submillimeter, and Far-Infrared Detectors and Instrumentation
  for Astronomy IX}}},\ Vol.\ \bibinfo {volume} {10708}\ (\bibinfo
  {organization} {Proc. SPIE},\ \bibinfo {year} {2018})\ p.\ \bibinfo {pages}
  {107081W}\BibitemShut {NoStop}%
\bibitem [{SMF()}]{SMF}%
  \BibitemOpen
  \href@noop {} {\enquote {\bibinfo {title} {{SMF - Special Metals
  Fabrication}},}\ }\bibinfo {howpublished} {{www.special-metals.co.uk}},\
  \bibinfo {note} {accessed: 02-09-2019}\BibitemShut {NoStop}%
\bibitem [{Mag()}]{MagneticShields}%
  \BibitemOpen
  \href@noop {} {\enquote {\bibinfo {title} {{Magnetic Shields}},}\ }\bibinfo
  {howpublished} {{www.magneticshields.co.uk}},\ \bibinfo {note} {accessed:
  02-09-2019}\BibitemShut {NoStop}%
\bibitem [{\citenamefont {Glowacka}\ \emph {et~al.}(2012)\citenamefont
  {Glowacka}, \citenamefont {Crane}, \citenamefont {Goldie},\ and\
  \citenamefont {Withington}}]{glowacka2012fabrication}%
  \BibitemOpen
  \bibfield  {author} {\bibinfo {author} {\bibfnamefont {D.}~\bibnamefont
  {Glowacka}}, \bibinfo {author} {\bibfnamefont {M.}~\bibnamefont {Crane}},
  \bibinfo {author} {\bibfnamefont {D.}~\bibnamefont {Goldie}}, \ and\ \bibinfo
  {author} {\bibfnamefont {S.}~\bibnamefont {Withington}},\ }\href@noop {}
  {\bibfield  {journal} {\bibinfo  {journal} {J. Low Temp. Phys.}\ }\textbf
  {\bibinfo {volume} {167}},\ \bibinfo {pages} {516} (\bibinfo {year}
  {2012})}\BibitemShut {NoStop}%
\bibitem [{\citenamefont {Drung}\ \emph {et~al.}(2007)\citenamefont {Drung},
  \citenamefont {Abmann}, \citenamefont {Beyer}, \citenamefont {Kirste},
  \citenamefont {Peters}, \citenamefont {Ruede},\ and\ \citenamefont
  {Schurig}}]{drung2007highly}%
  \BibitemOpen
  \bibfield  {author} {\bibinfo {author} {\bibfnamefont {D.}~\bibnamefont
  {Drung}}, \bibinfo {author} {\bibfnamefont {C.}~\bibnamefont {Abmann}},
  \bibinfo {author} {\bibfnamefont {J.}~\bibnamefont {Beyer}}, \bibinfo
  {author} {\bibfnamefont {A.}~\bibnamefont {Kirste}}, \bibinfo {author}
  {\bibfnamefont {M.}~\bibnamefont {Peters}}, \bibinfo {author} {\bibfnamefont
  {F.}~\bibnamefont {Ruede}}, \ and\ \bibinfo {author} {\bibfnamefont
  {T.}~\bibnamefont {Schurig}},\ }\href@noop {} {\bibfield  {journal} {\bibinfo
   {journal} {IEEE Transactions on Applied Superconductivity}\ }\textbf
  {\bibinfo {volume} {17}},\ \bibinfo {pages} {699} (\bibinfo {year}
  {2007})}\BibitemShut {NoStop}%
\bibitem [{\citenamefont {Lindeman}\ \emph {et~al.}(2004)\citenamefont
  {Lindeman}, \citenamefont {Bandler}, \citenamefont {Brekosky}, \citenamefont
  {Chervenak}, \citenamefont {Figueroa-Feliciano}, \citenamefont {Finkbeiner},
  \citenamefont {Li},\ and\ \citenamefont {Kilbourne}}]{lindeman2004impedance}%
  \BibitemOpen
  \bibfield  {author} {\bibinfo {author} {\bibfnamefont {M.~A.}\ \bibnamefont
  {Lindeman}}, \bibinfo {author} {\bibfnamefont {S.}~\bibnamefont {Bandler}},
  \bibinfo {author} {\bibfnamefont {R.~P.}\ \bibnamefont {Brekosky}}, \bibinfo
  {author} {\bibfnamefont {J.~A.}\ \bibnamefont {Chervenak}}, \bibinfo {author}
  {\bibfnamefont {E.}~\bibnamefont {Figueroa-Feliciano}}, \bibinfo {author}
  {\bibfnamefont {F.~M.}\ \bibnamefont {Finkbeiner}}, \bibinfo {author}
  {\bibfnamefont {M.~J.}\ \bibnamefont {Li}}, \ and\ \bibinfo {author}
  {\bibfnamefont {C.~A.}\ \bibnamefont {Kilbourne}},\ }\href@noop {} {\bibfield
   {journal} {\bibinfo  {journal} {Rev. Sci. Instrum}\ }\textbf {\bibinfo
  {volume} {75}},\ \bibinfo {pages} {1283} (\bibinfo {year}
  {2004})}\BibitemShut {NoStop}%
\bibitem [{\citenamefont {Rostem}, \citenamefont {Withington},\ and\
  \citenamefont {Goldie}(2007)}]{rostem2007multitone}%
  \BibitemOpen
  \bibfield  {author} {\bibinfo {author} {\bibfnamefont {K.}~\bibnamefont
  {Rostem}}, \bibinfo {author} {\bibfnamefont {S.}~\bibnamefont {Withington}},
  \ and\ \bibinfo {author} {\bibfnamefont {D.}~\bibnamefont {Goldie}},\
  }\href@noop {} {\bibfield  {journal} {\bibinfo  {journal} {J. Appl Phys}\
  }\textbf {\bibinfo {volume} {102}},\ \bibinfo {pages} {034511} (\bibinfo
  {year} {2007})}\BibitemShut {NoStop}%
\bibitem [{\citenamefont {Goldie}\ \emph {et~al.}(2009)\citenamefont {Goldie},
  \citenamefont {Audley}, \citenamefont {Glowacka}, \citenamefont {Tsaneva},\
  and\ \citenamefont {Withington}}]{goldie2009thermal}%
  \BibitemOpen
  \bibfield  {author} {\bibinfo {author} {\bibfnamefont {D.}~\bibnamefont
  {Goldie}}, \bibinfo {author} {\bibfnamefont {M.}~\bibnamefont {Audley}},
  \bibinfo {author} {\bibfnamefont {D.}~\bibnamefont {Glowacka}}, \bibinfo
  {author} {\bibfnamefont {V.}~\bibnamefont {Tsaneva}}, \ and\ \bibinfo
  {author} {\bibfnamefont {S.}~\bibnamefont {Withington}},\ }\href@noop {}
  {\bibfield  {journal} {\bibinfo  {journal} {J. Appl. Phys.}\ }\textbf
  {\bibinfo {volume} {105}},\ \bibinfo {pages} {074512} (\bibinfo {year}
  {2009})}\BibitemShut {NoStop}%
\end{thebibliography}%

\end{document}